%% file: VLM_paper_AIware2026_cameraReady.tex
\documentclass[sigconf]{acmart}
\AtBeginDocument{%
}

\usepackage{paralist}
\usepackage{stfloats}
\usepackage{xspace}
\newcommand{\Atwelve}{\ensuremath{\hat{A}}\textsubscript{12}\xspace}

\newcommand{\neglibleStrength}{\texttt{negligible}\xspace}
\newcommand{\smallStrength}{\texttt{small}\xspace}
\newcommand{\mediumStrength}{\texttt{medium}\xspace}
\newcommand{\largeStrength}{\texttt{large}\xspace}

\setcopyright{acmlicensed}
\copyrightyear{2026}
\acmYear{2026}
\acmDOI{XXXXXXX.XXXXXXX}
\acmConference[AIware 2026]{3rd ACM International Conference on AI-powered Software}{July 06--07, 2026}{Montreal, Canada}
\acmISBN{978-1-4503-XXXX-X/2018/06}

\newcommand{\approach}{\ensuremath{\mathtt{VISOR}}\xspace}
\newcommand\pimodel{$\pi_0$\xspace} 

\newcommand{\geminiLong}{\ensuremath{\mathtt{Gemini\text{-}2.5\text{-}Flash}}\xspace}
\newcommand{\gemini}{\ensuremath{\mathtt{Gemini}}\xspace}
\newcommand{\openaiLong}{\ensuremath{\mathtt{GPT\text{-}4.1}}\xspace}
\newcommand{\openai}{\ensuremath{\mathtt{GPT}}\xspace}

\newcommand{\LLaVANext}{\ensuremath{\mathtt{LLaVA\text{-}Next}}\xspace}
\newcommand{\SmolVLMTwo}{\ensuremath{\mathtt{SmolVLM2}}\xspace}
\newcommand{\Qwen}{\ensuremath{\mathtt{Qwen}}\xspace}
\newcommand{\Eagle}{\ensuremath{\mathtt{Eagle}}\xspace}

\newcommand{\movenear}{\ensuremath{\mathtt{Move Near}}\xspace}
\newcommand{\pickup}{\ensuremath{\mathtt{Pick Up}}\xspace}
\newcommand{\putin}{\ensuremath{\mathtt{Put In}}\xspace}
\newcommand{\puton}{\ensuremath{\mathtt{Put On}}\xspace}

\usepackage[most]{tcolorbox}
\usepackage[table]{xcolor}
\usepackage{multirow}
\usepackage{placeins}
\newtcolorbox{custombox}[1]{
colback=gray!10,
breakable,
colframe=gray!20,
left=0.5mm,
right=0.5mm,
top=0.5mm,
bottom=0.5mm,
fonttitle=\bfseries,
arc=1mm,
leftrule=0mm,
rightrule=0mm,
toprule=0mm,
bottomrule=0mm,
notitle,
before=\par\smallskip\noindent,
before upper={\textbf{#1: } },
}
\usepackage{enumitem}
\usepackage{listings}

\lstdefinelanguage{json}{
basicstyle=\ttfamily\footnotesize,
showstringspaces=false,
breaklines=true,
frame=none
}
\definecolor{ctxblue}{HTML}{D9E8FF}
\definecolor{inpgreen}{HTML}{DFF2DF}
\definecolor{reqpink}{HTML}{FFD9D9}
\definecolor{instorange}{HTML}{FFE6C7}
\definecolor{boxline}{HTML}{222222}

\newtcolorbox{promptsegment}[2][]{%
enhanced,
colframe=boxline,
colback=#2,
boxrule=0.5pt,
arc=2mm,
left=2mm,right=2mm,top=1.5mm,bottom=1.5mm,
before skip=1.5mm, after skip=1.5mm,
fonttitle=\bfseries\footnotesize,
title=#1
}

\newtcolorbox{promptcontainer}[1][]{%
enhanced,
colframe=boxline,
colback=white,
boxrule=0.9pt,
arc=2mm,
left=2mm,right=2mm,top=2mm,bottom=2mm,
#1
}

\usepackage[most]{tcolorbox}
\usepackage{tikz}
\usepackage{xcolor}
\usepackage{enumitem}
\usepackage{listings}
\usetikzlibrary{calc,decorations.pathreplacing}

\definecolor{ctxblue}{HTML}{D9E8FF}
\definecolor{inpgreen}{HTML}{DFF2DF}
\definecolor{reqpink}{HTML}{FFD9D9}
\definecolor{instorange}{HTML}{FFE6C7}
\definecolor{boxline}{HTML}{222222}

\newtcolorbox{promptbig}{%
enhanced,
colframe=boxline,
colback=white,
boxrule=0.9pt,
arc=2mm,
left=2mm,right=2mm,top=2mm,bottom=2mm
}

\newtcolorbox{bandbox}[2]{%
enhanced,
colframe=boxline,
colback=#2,
boxrule=0.6pt,
arc=1.6mm,
left=2mm,right=2mm,top=1.2mm,bottom=1.2mm,
before skip=0mm, after skip=0mm
}

\usetikzlibrary{positioning,fit,calc,decorations.pathreplacing}

\usetikzlibrary{positioning,fit,calc,decorations.pathreplacing}

\definecolor{rulespurple}{HTML}{E8E0FF}

\newtcolorbox{promptframe}[1][]{%
enhanced,
colframe=boxline,
colback=white,
boxrule=0.5pt,
arc=1mm,
left=0mm,right=0mm,top=0mm,bottom=0mm,
width=1\columnwidth,
#1
}

\newtcolorbox{promptband}[2][]{%
enhanced,
colback=#2,
colframe=#2,
boxrule=0pt,
arc=0mm,
left=0mm,right=0mm,top=0mm,bottom=0mm,
before skip=0mm, after skip=0mm,
#1
}

\begin{document}

\title[VISOR: A Vision-Language Model-based Test Oracle for Testing Robots]{VISOR: A Vision-Language Model-based Test Oracle for\\Testing Robots}

\author{Prasun Saurabh}
\email{prasun@simula.no}
\orcid{0009-0000-7139-1043}
\affiliation{%
  \institution{Simula Research Laboratory and Oslo Metropolitan University}
\city{Oslo}
  \country{Norway}
}

\author{Pablo Valle}
\email{pvalle@mondragon.edu}
\orcid{0000-0002-0588-316X}
\affiliation{%
\institution{Mondragon University}
\city{Mondragon}
\country{Spain}
}

\author{Aitor Arrieta}
\email{aarrieta@mondragon.edu}
\orcid{0000-0001-7507-5080}
\affiliation{%
\institution{Mondragon University}
\city{Mondragon}
\country{Spain}
}

\author{Shaukat Ali}
\email{shaukat@simula.no}
\orcid{0000-0002-9979-3519}
\affiliation{%
\institution{Simula Research Laboratory}
\city{Oslo}
\country{Norway}
}

\author{Paolo Arcaini}
\email{arcaini@nii.ac.jp}
\orcid{0000-0002-6253-4062}
\affiliation{%
\institution{National Institute of Informatics}
\city{Tokyo}
\country{Japan}
}


\begin{abstract}
Testing robots requires assessing whether they perform their intended tasks correctly, dependably,  and with high quality, a challenge known as the test oracle problem in software testing. Traditionally, this assessment relies on task-specific symbolic oracles for task correctness and on human manual evaluation of robot behavior, which is time-consuming, subjective, and error-prone. To address this, we propose \approach, a Vision-Language Model (VLM)–based approach for automated test oracle assessment that eliminates the need of expensive human evaluations. \approach performs automated evaluation of task correctness and quality, addressing the limitations of existing symbolic test oracles, which are task-specific and provide pass/fail judgments without explicitly quantifying task quality. Given the inherent uncertainty in VLMs, \approach also explicitly quantifies its own uncertainty during test assessments. We evaluated \approach using two VLMs, i.e., \openai and \gemini, across four robotic tasks on over 1,000 videos. Results show that \gemini achieves higher recall while \openai achieves higher precision. However, both models show low correlation between uncertainty and correctness, which prevents using uncertainty as a correctness predictor.
\end{abstract}



\maketitle
\section{Introduction}\label{sec:intro}

Robots are increasingly being deployed across a wide range of applications in complex and dynamic environments, where they are expected to perform tasks correctly, meet performance and safety requirements, and avoid harming users and other entities~\cite{asif2025rapid, rodriguez2021human,pantalone2021robot,li2024robonurse}. Moreover, robots are becoming increasingly autonomous~\cite{nvidia2025gr00tn1openfoundation} and are being deployed in safety and mission-critical applications, making their dependability and high-quality behavior extremely important.

This calls for systematic and automated testing of robots to ensure not only successful task completion but also adequate task quality, as defined by user criteria. In software testing, this problem is referred to as the {\it test oracle problem}~\cite{oracleProblemTSE2014}, which determines whether software meets its intended behavior. In robotics, this means assessing whether a robot performs its intended task correctly and to an acceptable standard. Symbolic test oracles, which are commonly employed in this context~\cite{wang2025vlatest, zhang2025vlabench, peng2025nebula}, are typically task-specific, lack reusability across different tasks, and provide only binary pass/fail judgments, making them unable to automatically assess task quality. Task quality assessment often requires human involvement for validation~\cite{VLAvalleArXiv2025}, as humans visually inspect robot executions to judge task quality. This practice is time-consuming, error-prone, difficult to scale, and subjective, as different human testers may perceive task quality differently for the same test. As a result, testing robots at scale remains difficult without automation.

Vision-Language Models (VLMs) are large models capable of understanding and reasoning about images, videos, and text. They are increasingly being used for solving problems in several domains, including robotics~\cite{Chen_2024_CVPR,duan2024aha, chen2023autotamp,VLMSocialNav}. In this paper, we use VLMs to automate test oracle assessment for robots, by proposing a kind of VLM-as-a-judge approach called \approach. The approach automatically assesses task correctness and quality from robot execution videos. Moreover, given the inherent uncertainty of VLMs, \approach explicitly quantifies this uncertainty, providing a measure of trustworthiness for each evaluation to determine whether the VLM's assessment can be relied upon or should be treated with caution.

We conducted a large-scale empirical analysis of \approach using two VLMs, namely \openai and \gemini, on a dataset of over 1,000 videos spanning four different tasks of robotic arms, in which robots either successfully completed the tasks or failed to do so. We also define a new distance metric for the quality analysis that estimates the severity of misclassification in quality assessment. Experimental results reveal that \gemini excels at recall, while \openai produces more precise predictions aligned with ground truth. Both models exhibit low uncertainty and stable performance across different runs.

\section{Background}\label{sec:background}
In this section, we present the foundational concepts and background related to Vision-Language Models and Robotic Systems.

\subsection{Vision-Language Models} \label{subsec:VLMs}
VLMs represent a major advance in foundational multimodal artificial intelligence, enabling integrated perception, understanding, and reasoning across visual inputs (such as images or videos) and natural language~\cite{radford2021learning, wang2022git}. They are typically pre-trained on massive internet-scale datasets pairing images with descriptive captions, text, or other aligned multimodal pairs, fostering deep semantic alignments between visual content and linguistic representations~\cite{liu2023visual}. Recent architectural advances have further enhanced their generality, efficiency, and scalability. Modern VLMs often integrate frozen high-capacity vision encoders (e.g., CLIP~\cite{radford2021learning} and SigLIP~\cite{tschannen2025siglip}), powerful large language model backbones~\cite{kawaharazuka2025vision}, and lightweight modality-fusion components such as projectors or adapters~\cite{liu2023visual}. These advances support a wide variety of real-world applications~\cite{zhang2024vision}, such as answering questions about medical images or documents~\cite{hartsock2024vision} or content generation through summarization of infographics~\cite{baechler2024screenai}.

In robotics, VLMs serve as a powerful perception and high-level reasoning backbone, enabling robots to interpret their surroundings and understand natural language instructions. This has lead the emergence of Vision-Language-Action (VLA) models~\cite{nvidia2025gr00tn1openfoundation,black2024pi0visionlanguageactionflowmodel}, which extend VLMs by incorporating dedicated action-generation heads to control the robot. VLAs directly map combined visual and linguistic inputs to low-level robot control signals for embodied tasks~\cite{zitkovich2023rt, kim2024openvla, kawaharazuka2025vision}.

\vspace{2pt}
\subsection{Robotic Systems}\label{subsec:VLA-robots}
Cyber-Physical Systems (CPSs) integrate computation, communication, and physical processes to sense, reason about, and act upon the real world~\cite{baheti2011cyber, derler2011modeling}. They operate in safety-critical domains such as autonomous vehicles~\cite{kato2018autoware}, surgical and assistive medical robotics~\cite{dey2018medical}, and advanced manufacturing~\cite{monostori2016cyber}. Robotic systems represent a core subclass of CPSs, where embodied agents must perceive dynamic environments, make decisions, and execute physical actions with precision and safety.

In this work, we focus on robotic manipulation systems, primarily articulated robotic arms fitted with end-effectors such as parallel-jaw grippers. They perform a wide variety of tasks, like grasping and placing objects. Reliable operation depends on tightly integrated components: motor controllers for actuation, proprioceptive sensors for internal state estimation, exteroceptive sensors for environmental awareness, and planning-control algorithms that generate feasible, collision-free actions adapted to task goals. The recent irruption of VLA models~\cite{nvidia2025gr00tn1openfoundation,black2024pi0visionlanguageactionflowmodel} significantly changed robotic manipulation systems. Traditional pipelines that handle perception, planning, and control separately can now be replaced by a single end-to-end learned model. A VLA-enabled robot can now directly interpret natural language instructions such as \textit{``pick up the water bottle''}, while processing live visual observations to generate executable actions.



\section{\approach}\label{sec:approach}

\begin{figure*}[!t]
\centering
\includegraphics[width=0.9\linewidth]{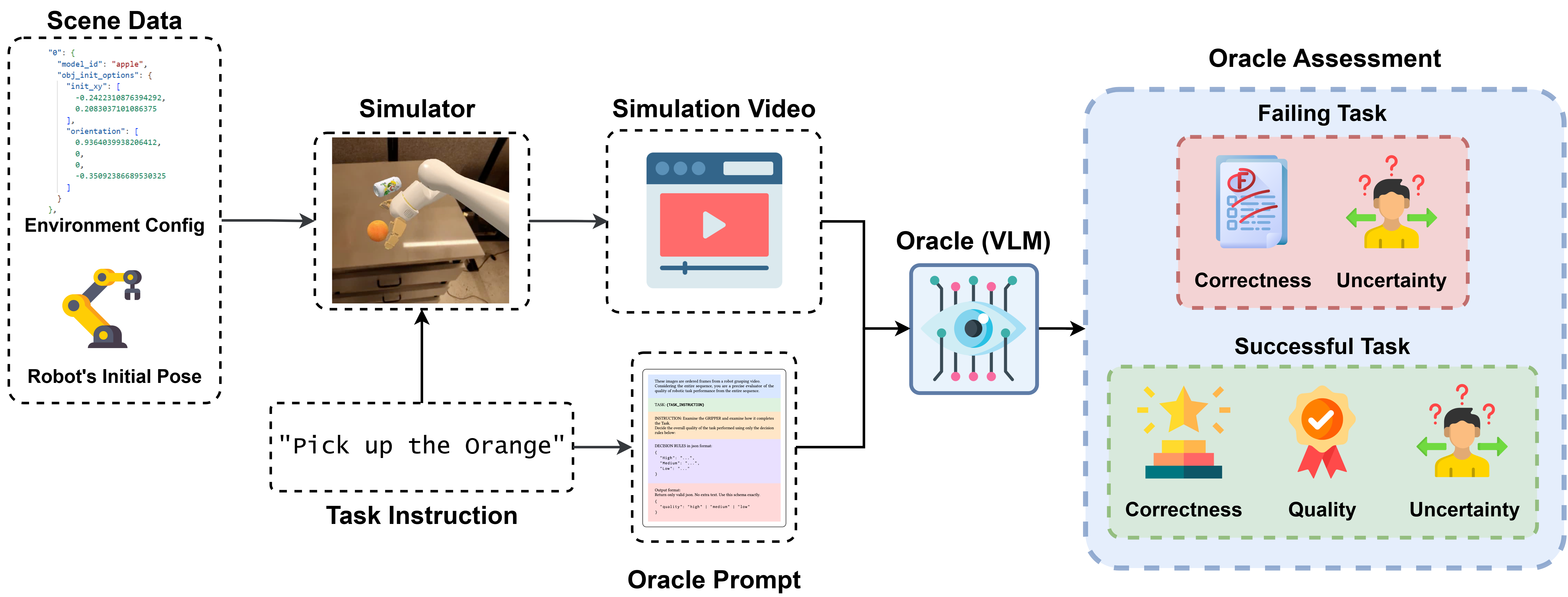}
\caption{\approach{} -- In \textit{Failing Task}, Correctness refers to failure; in \textit{Successful Task}, Correctness refers to success.}
\Description{Overview of the approach}
\label{fig:experiment_process}
\end{figure*}

Fig.~\ref{fig:experiment_process} presents an overview of the proposed approach \approach.
We use a VLM as a test oracle to analyze videos of robots performing various tasks and assess their correctness and quality, rather than relying on manual human analysis of robots or of their videos for this purpose. In parallel, we explicitly quantify the VLM's uncertainty when performing these tasks, enabling interpretation of the reliability of automated task correctness and quality assessments.

In the context of this paper, such videos are produced by a simulator that takes as input initial scene data, such as the position (i.e., $x$, $y$, $z$) and rotation in quaternions (i.e., $q1$, $q2$, $q3$, $q4$) of each object in the scene. In addition to this, each object in the scene is represented with a $\mathit{model_{id}}$, which is used by the simulator to determine which is the object to be displayed since, for the same object (e.g., a can), different textures might be used (e.g., cola, fanta, 7up, \ldots). All this represents environmental configuration and together with the robot's initial pose, along with task instructions (e.g., ``Pick up the orange''), constitute the inputs to the simulator. Next, the simulator, along with a control policy executes the task and generates a corresponding video, which is then assessed by a VLM to determine task correctness and quality. Such an assessment can determine whether a task has failed or succeeded. In the latter case, the VLM further assesses {\it task quality}, classifying it as {\it High}, {\it Medium}, or {\it Low}. Moreover, we measure the uncertainty of the task correctness prediction for each video by analyzing how confident it is in determining whether a task has been performed with success or fail. Similarly, for task quality assessment, we measure the uncertainty when distinguishing among \textit{High}, \textit{Medium}, and \textit{Low} quality prediction. The specific metrics used to quantify uncertainty are described in Sect.~\ref{sec:uncertainty_metrics}.

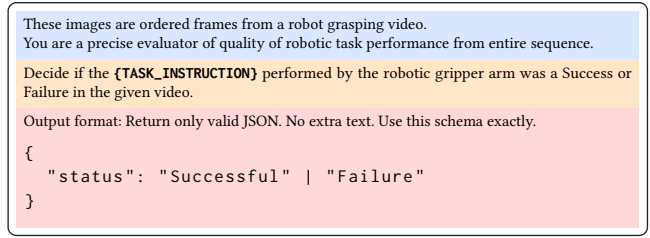
\begin{figure}[!t]
\input{images/prompt_binary}
\caption{Prompt template for task correctness assessment.}
\Description{Prompt template for task correctness assessment.}
\label{fig:prompt-template-binary}
\end{figure}
\begin{figure}[!t]
\input{images/prompt_multi}
\caption{Prompt template for task quality assessment.}
\Description{Prompt template for task quality assessment.}
\label{fig:prompt-template-multiclass}
\end{figure}
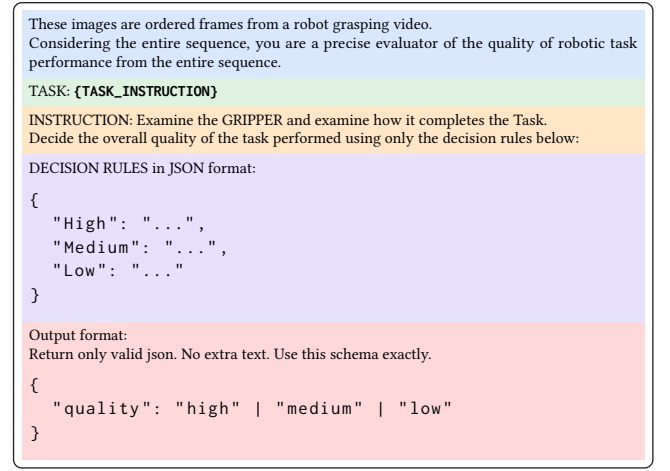

A VLM requires a well-designed prompt to assess task correctness and quality of its execution~\cite{Sahoo2024PromptEngineeringSurvey}. To this end, we first made a small-scale evaluation of several prompting styles, including zero-shot and few-shot. Our preliminary experiments and analysis across these prompting styles revealed a substantial increase in the number of input tokens for few-shot prompting (i.e., 64K tokens) compared to zero-shot (i.e., 7K tokens), leading to significantly higher inference time and computational cost. As a result, few-shot prompting proved to be too expensive relative to the zero-shot approach. Therefore, we decided to use the zero-shot prompting style. 
We then iteratively refined the prompt to achieve more reliable outputs. Consequently, we designed two prompts: one for task correctness assessment (Fig.~\ref{fig:prompt-template-binary}) and the other for test quality assessment (Fig.~\ref{fig:prompt-template-multiclass}).

\vspace{-4pt}

Fig.~\ref{fig:prompt-template-binary} shows the prompt template for task correctness assessment. The first part of the prompt assigns the VLM the role of a test oracle and provides context by specifying that video frames are presented as an ordered sequence. It also includes task instructions for each video, e.g., ``Pick up the Orange'' specified in the \textit{TASK\_INSTRUCTION} variable, along with details on what the VLM must observe and classify. Finally, the prompt enforces output requirements, requiring the VLM to generate a JSON-only schema with a binary label (i.e., Successful or Failure).

Fig.~\ref{fig:prompt-template-multiclass} shows the prompt template we used for task quality assessment.
Like the task correctness prompt, it assigns the VLM the role of a test oracle and provides contextual information. Next, it provides a task instruction (e.g., ``Pick up the Orange''), specified in the \textit{TASK\_INSTRUCTION} variable, along with details on the task quality to be assessed. It also defines predefined decision rules, which are the ones Valle et. al.~\cite{VLAvalleArXiv2025} used to guide the human assessment in their experiment, that map task execution to three quality levels (i.e., {\it High}, {\it Medium}, or {\it Low}). Finally, it requires the VLM to produce a strict JSON-only output using a fixed schema.

\section{Empirical Evaluation}\label{sec:empiricalevaluation}
We conducted an empirical study to assess the effectiveness of VLMs in evaluating the correctness and quality of robotic tasks. This section presents our research questions and details of the experimental setup used for the evaluation. All scripts, benchmarks, and results are available at~\cite{SupplementaryMaterialAIWARE2026}.

\subsection{Research Questions (RQs)}\label{subsec:rqs}

Our overall objective is to evaluate the effectiveness of VLMs as automated test oracles for robotic systems by assessing their ability to determine task success and task quality. Based on this objective, we aim to address the following research questions (RQs):
\begin{compactitem}
\item[\textbf{RQ1}] \textit{How effective are VLMs as automated test oracles for robotic task execution?}\\
This research question investigates whether VLMs can effectively serve as test oracles for assessing task success or failure (RQ1.a) and task quality (RQ1.b). To this end, we further subdivide this RQ into two sub-questions.
\begin{compactitem}
\item[\textbf{RQ1.a}] \textit{How accurately can VLMs determine task success or failure from robot execution videos for different tasks?}\\
This RQ evaluates the performance of VLMs as classifiers for the task correctness by labeling robot execution videos as either successful or failed.
\item[\textbf{RQ1.b}] 
\textit{How effectively can VLMs assess the quality of successfully completed robotic tasks?}\\
This RQ evaluates the performance of VLMs in classifying the quality of successful executions using three categories: {\it High}, {\it Medium}, or {\it Low}. This results in a multi-class classification problem to be solved by VLMs.
\end{compactitem}
\item[\textbf{RQ2}] \textit{How uncertain are VLMs in their assessments of task correctness and task quality?}\\
This RQ investigates the uncertainty of VLM predictions, that is, how confident the VLMs are for both task correctness and task quality evaluations.
\item[\textbf{RQ3}] \textit{To what extent does VLM uncertainty correlate with test oracle decisions for task correctness and quality assessment?}\\
Studying this relationship is important to determine whether uncertainty can serve as a reliable indicator of assessments of task correctness and task quality. It also helps identify which uncertainty metrics are most suitable for evaluating the trustworthiness of VLMs for test oracle assessment.
\end{compactitem}


\subsection{Evaluation Dataset} \label{subsec:casestudies}
For our evaluation, as a ground truth dataset, we used the videos provided by Valle et al.~\cite{VLAvalleArXiv2025}, which comprise four tasks (i.e., \movenear, \pickup, \putin, and \puton) across three Visual Language Action (VLA) models (i.e., OpenVLA~\cite{kim2024openvla}, \pimodel~\cite{black2024pi0visionlanguageactionflowmodel}, and SpatialVLA~\cite{qu2025spatialvla}). The evaluation has 500 videos per task and VLA model, forming an unbalanced dataset of successful and failing task executions. Moreover, among successful executions, the distribution of task quality (i.e., {\it High}, {\it Medium}, {\it Low}) is also uneven. To address this, we curated a balanced subset for our evaluation as follows.

For each task, we first determined the maximum number of successful videos available at each task quality, across all VLAs. For example, for \pickup, the numbers of successful videos are 201 ({\it High}), 66 ({\it Medium}), and 117 ({\it Low}); so, for this task, we set the maximum number of videos for each quality level as 66. We then selected 66 videos for each quality, attempting to balance across VLAs. Ideally, this would be 22 videos per VLA model, per quality level. However, for models with fewer than 22 videos, we included all their videos and distributed the remaining videos among the other models. For instance, for {\it High} quality videos for \pickup, we selected the 18 videos from OpenVLA, and we balanced the remaining ones by selecting 24 each from \pimodel and SpatialVLA.

Applying this procedure across all tasks, we obtained 198 videos for \pickup, 141 for \movenear, 66 for \putin, and 111 for \puton, totaling 516 successful videos. To balance the dataset, we included an equal number of failing videos per task, that is 198, 141, 66, and 111, respectively. Since sufficient failing videos were available, this selection could be fully balanced across models. The resulting evaluation dataset comprised 1,032 videos.

\subsection{Selected Vision-Language Models} \label{subsec:selectedvlms}

We conducted a small exploratory experiment to select the VLMs for our study. We assessed six VLMs from open-source and proprietary categories: \openaiLong~\cite{gpt4Arxiv}, \geminiLong~\cite{comanici2025gemini}, \LLaVANext~\cite{LLaVANextarxiv}, \SmolVLMTwo~\cite{SmolVLMArxiv}, \Qwen~\cite{qwenArxiv}, and \Eagle~\cite{EagleVLM}. For a fair comparison, we used the same settings across all VLMs, including a temperature of 0 to minimize output stochasticity. We ran the experiment five times, selecting six video files from each of the four tasks.

For task correctness assessment \openaiLong and \geminiLong produced more consistent and task-sensitive results across the four tasks. In contrast, the open-source models (\LLaVANext, \SmolVLMTwo, \Qwen, and \Eagle) exhibited identical scores (i.e., Precision=0.500, Recall=1.000, F1=0.667) across all tasks, indicating a behavior that does not reliably identify the outcome. We observed that all open-source VLMs predicted all videos as ``Successful'' in the correctness assessment, never resulting in a ``False Negative''. Consequently, the higher recall results from this consistent prediction behavior rather than evidence of strong task understanding, and it does not do justice to the skewed outcome. Because on this lack of task sensitivity, we excluded these open-source models from the full experiments, and we only used \openaiLong and \geminiLong (called \openai and \gemini in the following).





\subsection{Evaluation Metrics} \label{subsec:evaluatiometrics}

\noindent{\bf Evaluation Metrics for RQ1.a.}
We apply usual binary classification metrics, i.e., {\it precision}, {\it recall}, and {\it F1}, to assess the performance of the selected VLMs per task. Since each experiment is repeated 10 times, we report the mean and standard deviation of each metric to show variability in effectiveness across runs. To assess the statistical significance, we apply the Mann-Whitney U test for each metric and task. A $p$-value less than 0.05 indicates statistically significant differences. In addition, we report the Vargha-Delaney effect size \Atwelve to quantify the magnitude of the observed differences.

For each metric (i.e., precision, recall, or F1), \Atwelve estimates the probability that one VLM (V1) yields higher values than another VLM (V2). \(\Atwelve = 0.5\) indicates no difference, while values greater than 0.5 indicate that V1 is more likely to achieve higher metric values than V2. We interpret the results following the literature~\cite{mangiafico2016summary}: 1) \neglibleStrength\: \(\Atwelve \in (0.44, 0.56)\), 2) \smallStrength\ for \(\Atwelve \in (0.34, 0.44]\) or \(\Atwelve \in [0.56, 0.64)\), 3) \mediumStrength\ for \(\Atwelve \in (0.29, 0.34]\) or \(\Atwelve \in [0.64, 0.71)\), and 4) \largeStrength\ for \(\Atwelve \in [0, 0.29]\) or \(\Atwelve \in [0.71, 1]\).

\vspace{3pt}
\noindent{\bf Evaluation Metrics for RQ1.b.} We evaluate the VLMs performance on the multi-class task of assessing task quality with multi-class classification metrics, i.e., Precision-micro, Precision-macro, Recall-micro, Recall-macro, F1-micro, and F1-macro. The macro versions of these metrics treat each class equally, providing an overall measure of effectiveness across all quality categories (e.g., {\it High}, {\it Medium}, {\it Low}), while the micro versions aggregate contributions from all samples, giving greater influence to classes with more instances. Using both versions of the metrics allows us to evaluate the overall performance of the VLMs across classes and their performance under class imbalance. 

In addition, we define a {\it distance} metric $d \in \{0,\ldots, 5\}$ to quantify the discrepancy between a predicted task outcome and the ground truth by considering both task correctness and task quality. Let $c \in \{\textit{pass}, \textit{fail}\}$ denote the ground-truth correctness and $\hat{c}$ the predicted correctness. The correctness component of the distance is defined as $d_c = 0$ if $\hat{c} = c$, and $d_c = 2$ otherwise. Task quality is assumed to be ordinal with levels $q \in \{\textit{Failure}, \textit{Low}, \textit{Medium}, \textit{High}\}$, which we map to numerical values $\{0, 1, 2, 3\}$. Let $\hat{q}$ denote the predicted quality level; the quality component of the distance is then $d_q = |\hat{q} - q|$. The final distance is computed as the sum $d = d_c + d_q$, which ranges from 0 for a perfect prediction to 5 for a maximally incorrect prediction.

To assess statistical significance between the two selected VLMs for each metric, we apply the Mann–Whitney U test and report the Vargha–Delaney \Atwelve effect size, as in RQ1.a. For the distance metric, we compute, for each task, run and VLM, the mean distance across all videos, where lower values indicate better performance. For all other metrics, higher values correspond to better performance.

\vspace{3pt}
\noindent{\bf Evaluation Metrics for RQ2.}\label{sec:uncertainty_metrics} We analyze the uncertainty of the models in order to determine how reliable their answer is. We selected one uncertainty metric derived from model outputs, i.e., Entropy~\cite{bishop2006}, which quantifies the dispersion of the predicted probability distribution over output classes. We also compute commonly used confidence-based metrics, i.e., Maximum Softmax Probability (MSP) and DeepGini~\cite{feng2020deepgini}, defined as the difference between the highest and second-highest predicted class probabilities. Larger margins indicate more confident predictions, whereas smaller margins suggest higher uncertainty.

\vspace{3pt}
\noindent{\bf Evaluation Metrics for RQ3.}
In RQ3, we focus on finding the relationship between uncertainty and performance of the models. Specifically, we analyze the correlation between the proposed distance metric and the uncertainty measures from RQ2. By focusing on distance rather than accuracy alone, we account for the severity of prediction errors and examine whether higher uncertainty is associated with larger deviations from the ground truth. We perform Spearman's correlation test to quantify the strength and direction of this relationship, thereby assessing whether uncertainty estimates can serve as reliable indicators of prediction quality. The value of the correlation coefficient ($\rho$) ranges from -1 to +1, where negative values indicate a negative correlation and positive values indicate a positive correlation. We also report the $p$-value, where a value lower than 0.05 indicates a statistically significant correlation.

\section{Analysis of the Results} \label{sec:analysisanddiscussion}

In this section, we present a detailed analysis of the results obtained for each of the research questions posed in this paper. Our goal is to evaluate the capabilities of VLMs as test oracles, highlighting both their strengths and limitations in assessing robotic task execution.

\subsection{RQ1: Effectiveness of VLMs as Test Oracles} \label{subsec:rq1results}

This research question assesses the effectiveness of VLMs as test oracles. Specifically, we aim to determine how well these models evaluate two key aspects of task performance:
\begin{inparaenum}[(i)]
\item {\it correctness}, i.e., whether the task was completed as intended, and
\item {\it quality}, i.e., how well the task was executed according to defined standards.
\end{inparaenum}

\subsubsection{RQ1a -- Effectiveness of VLMs as Test Oracle for Judging Task Correctness}
As shown in Table~\ref{tab:binary-summary}, for \movenear, \gemini achieves a precision of 0.561 and a high recall of 0.730, resulting in an overall F1 score of 0.635.
\begin{table}[!t]
\centering
\caption{RQ1.a -- Precision, recall, and F1 scores of \openai and \gemini for the four tasks for task correctness.}
\label{tab:binary-summary}
\input{tables/table_rq1a_recall_precision_f1}
\end{table}
This indicates that \gemini correctly classifies most test assessments but produces some false positives. In contrast, \openai exhibits high precision (0.844) but low recall (0.251), yielding a low F1 score of 0.386. While \openai makes very few false positives, it misses many correct test assessments. As shown in Table~\ref{tab:pairwise_mwu_metrics}, statistical tests confirm that \openai significantly outperforms \gemini in precision with a large effect size, whereas \gemini is significantly better than \openai for recall and F1, also with large effect sizes.
\begin{table}[!t]
\centering
\caption{RQ1.a -- Mann-Whitney U-Test and \Atwelve statistics to compare \openai vs \gemini in terms of task task correctness.}
\label{tab:pairwise_mwu_metrics}
\input{tables/table_rq1a_stats}
\end{table}
Generally, the standard deviations are very low (Table~\ref{tab:binary-summary}), indicating that the performance of the models across different runs is stable.

As shown in Table~\ref{tab:binary-summary}, for \pickup, \gemini obtained a precision of 0.609 and a high recall of 0.836, resulting in an F1 score of 0.704. This suggests that \gemini captures most of the correct test assessments, though it also produces a moderate number of false positives. In comparison, \openai demonstrates higher precision (i.e, 0.775) but lower recall (i.e., 0.541) than \gemini, yielding a lower F1 score of 0.637. Furthermore, as Table~\ref{tab:pairwise_mwu_metrics} depicts, the results are the same as for \movenear, i.e., \openai significantly outperforms \gemini in precision with a large effect size, whereas \gemini is significantly better in recall and F1 than \openai, also with large effect sizes. The low standard deviations (Table~\ref{tab:binary-summary}) indicate that overall both models' performance is consistent and stable across different runs.

Table~\ref{tab:binary-summary} shows that for \putin, \openai outperforms \gemini in precision (0.980 vs. 0.782), recall (0.739 vs. 0.651), and F1 score (0.842 vs. 0.711). As reported in Table~\ref{tab:pairwise_mwu_metrics}, these differences are statistically significant, with large effect sizes across all three metrics. For \puton, \openai achieves higher precision (0.959 vs. 0.674) and F1 score (0.839 vs. 0.759) than \gemini, but lower recall (0.746 vs. 0.869). Statistical tests in Table~\ref{tab:pairwise_mwu_metrics} confirm these differences, with all comparisons statistically significant and large effect sizes. All standard deviation values are low (Table~\ref{tab:binary-summary}), indicating stable performance across runs.

Based on these results, we conclude that across the four tasks, \gemini typically achieves higher recall, capturing more true test assessments, while \openai generally exhibits higher precision, producing fewer false positives. Thus, we suggest that using both models simultaneously may be beneficial to improve test assessment performance. Their outputs, for example, through a voting mechanism, could then be used to determine the final test assessment.

\vspace{-0.4em}
\subsubsection{RQ1.b -- Effectiveness of VLMs as Test Oracle for Judging Task Quality}
Table~\ref{tab:multiclass-summary} summarizes the results of assessing VLMs for judgment test quality. 
\begin{table*}[!t]
\centering
\caption{RQ1.b -- Precision/recall/F1 (micro and macro) and distance of \openai and \gemini for the four tasks (multiclass task-quality). }
\label{tab:multiclass-summary}
\input{tables/table_rq1b_recall_precision_f1}
\end{table*}
Overall, precision, recall, and F1 (both micro and macro) are relatively low across tasks. For example, the best performance is observed for the \puton task using \openai, with a precision-micro of 0.483, recall-micro of 0.483, and F1-micro of 0.483. For the remaining tasks, and across both models, the other metric values are even lower, indicating that the exact multiclass judgment test assessment remains challenging for the evaluated VLMs. The main reason behind this challenge is that the current VLMs are limited in reasoning about spatial changes in the datasets. Moreover, the three quality classes \textit{High}, {\it Medium}, and {\it Low} often differ by only fine-grained execution. This is even more evident in the \movenear task, where detecting the spatial gap proved more difficult.

When comparing the two models (Table~\ref{tab:multiclass_pairwise}), for \pickup, \putin, and \puton, \openai outperforms \gemini on all metrics with large effect sizes, except for recall-macro on \putin where there are no significant differences.
\begin{table*}[t]
\centering
\caption{RQ1.b -- Mann-Whitney U-Test and \Atwelve statistics to compare \openai vs \gemini in terms of task quality.}
\label{tab:multiclass_pairwise}
\input{tables/table_rq1b_stats}
\end{table*}
In contrast, for the \movenear task, \gemini significantly outperformed \openai on precision-micro, recall-micro, recall-macro, and F1-micro, with large effect sizes. There were no significant differences between the models for precision-macro; however, \openai achieved significantly higher F1-macro than \gemini. 

However, when examining the distance metric, we can see that for \gemini, the lowest (best) distance is achieved on \putin (1.794), while the highest (worst) value is observed for \movenear (2.506). For \openai, the best performance is again obtained on \puton, with a substantially lower distance of 0.843, whereas the worst performance occurs on \movenear, with a distance of 1.856. This suggests that although the models often fail to predict the exact quality class, their predictions are generally close to the true quality levels. \openai exhibits significantly lower distances than \gemini, with large effect sizes, as shown in Table~\ref{tab:multiclass_pairwise}, suggesting that \openai produces judgment test quality assessments that are closer to the ground truth.

\begin{custombox}{Answer to RQ1}
Our results suggest complementary strengths between the two VLMs: \gemini excels at identifying the most relevant test assessments (high recall), whereas \openai produces more precise predictions closer to the ground truth.
\end{custombox}

\subsubsection{Qualitative analysis of misclassification}
We conducted a qualitative analysis to understand \approach's limitations in classifying task correctness and task quality. 
\vspace{-0.4em}

\paragraph{Task Correctness misclassification:} We observed three recurrent patterns in task-correctness misclassification:
\begin{enumerate}[leftmargin=*, itemsep=2pt, topsep=2pt]
    \item \textbf{Near-complete executions were classified as success,} and borderline cases, where the robotic arm almost placed the object onto the target but did not fully complete the task, were often still judged as successful.
    
    \item \textbf{The wrong object interaction was classified as success.} In some cases, the robotic arm interacted with the wrong object, but the VLM still predicted success. For example, the arm picked up the Coca-Cola can and placed it in the basket, although the task required placing the Pepsi can.
    
    \item \textbf{Task-like motion classified as success.} The robotic arm sometimes appeared to perform the intended motion without fully completing one part of the task, yet such executions were still classified as successful.
\end{enumerate}

These patterns were especially pronounced for \gemini, which misclassified failure videos as success more frequently than \openai, particularly for \textit{MoveNear} and \textit{PickUp}.

\paragraph{Task Quality misclassification:}For task-quality prediction, we observed three recurrent failure modes:
\begin{compactenum}[(1)]
    \item \textbf{Overlooking subtle execution defects:} cues such as hesitation, vibration, unstable grasp, slight collision, or brief loss of control were often missed by the VLM, thus affecting the task quality classification.
    
    \item \textbf{Ignoring mid-execution quality-relevant events:} the VLM sometimes emphasized the final state of the task while overlooking events occurring in the middle of the execution, which affected the overall quality.
    
    \item \textbf{Borderline and late-stage errors:} the VLM struggled with borderline cases and with issues arising near the end of the execution, such as dropping or hitting the object, often predicting a higher quality class than intended.
\end{compactenum}

\subsection{RQ2: Uncertainty Assessment} \label{subsec:rq2results}

Table~\ref{tab:rq2-binary-uncertainty} presents descriptive statistics for \gemini and \openai across the four tasks for the three uncertainty metrics.
\begin{table}[!t]
\centering
\caption{RQ2 -- DeepGini, Entropy, and MSP of \openai and \gemini for the four tasks for task correctness. $\uparrow$ means a high value represents low uncertainty and vice versa.}
\label{tab:rq2-binary-uncertainty}
\input{tables/table_rq2_binary_uncertainty}
\end{table}
The results show that for DeepGini and Entropy, the values are very close to zero, whereas for MSP, all values are near 1.0, indicating that both models exhibit extremely low uncertainty for all the tasks. Moreover, the very low standard deviations across all metrics indicate that both models are highly stable across runs. Similar patterns are observed for task quality assessment across all four tasks (Table~\ref{tab:multiclass-uncertainty}), where uncertainty remains low, and model uncertainty is consistently stable across runs (low standard deviation).
\begin{table}[!t]
\centering
\caption{RQ2 -- DeepGini, Entropy, and MSP of \openai and \gemini for the four tasks for task-quality assessment. $\uparrow$ means a high value represents low uncertainty and vice versa.}
\label{tab:multiclass-uncertainty}
\input{tables/table_rq2_multiclass_uncertainty}
\end{table}

The consistently low uncertainty values observed across all tasks and both models should be interpreted with caution. While low uncertainty may indicate confident predictions, it does not necessarily imply correctness, particularly for task quality assessment, where effectiveness metrics remain modest. This suggests that the VLMs are often highly confident even when their predictions deviate from the ground truth, pointing to potential overconfidence in their internal probability estimates. This behavior is especially evident when contrasting the low uncertainty values in Tables~\ref{tab:rq2-binary-uncertainty} and \ref{tab:multiclass-uncertainty} with the relatively high distance values reported for task quality assessment. From a test oracle perspective, this highlights an important limitation: uncertainty estimates alone are insufficient to guarantee correctness and should not be interpreted in isolation. Nevertheless, the high stability of uncertainty metrics across runs indicates that uncertainty estimation itself is reproducible and consistent, making it a viable signal to be combined with effectiveness metrics or post-hoc calibration techniques when assessing the reliability of VLM-based test oracles for robotic manipulation applications.

\begin{custombox}{Answer to RQ2}
Both models exhibit consistently low uncertainty and stable performance across tasks. However, low uncertainty does not imply correctness, and models may exhibit high confidence even when classifying incorrectly. Thus, uncertainty metrics should be considered with effectiveness metrics for reliability and correctness.
\end{custombox}

\subsection{RQ3: Correlation Correctness-Uncertainty} \label{subsec:rq3results}

Table~\ref{tab:spearman-distance-uncertainty} reports Spearman's rank correlations between the uncertainty metrics and distance.
\begin{table}[!t]
\centering
\caption{RQ3 -- Spearman correlation between distance and uncertainty metrics (DeepGini, Entropy, and MSP). Higher DeepGini and Entropy values indicate higher uncertainty, whereas higher MSP values indicate lower uncertainty.}
\label{tab:spearman-distance-uncertainty}
\input{tables/table_rq3_spearman}
\end{table}
For \gemini, DeepGini, and Entropy exhibit significant positive correlations ($p < 0.05$, $\rho > 0$), indicating that higher uncertainty measured by these metrics is associated with larger distances, while lower uncertainty corresponds to smaller distances. For MSP, the same pattern is observed across all four tasks; however, since MSP is interpreted inversely, the correlations are negative ($\rho < 0$). Overall, these results suggest that higher uncertainty in \gemini is generally associated with predictions that deviate more from the ground truth.

For \openai, similar patterns to \gemini are observed for \movenear, \pickup, and \puton across the three uncertainty metrics. However, for \movenear, the correlations are not statistically significant ($p > 0.05$). In contrast, for \putin, the direction of the correlation is reversed, indicating a negative association between uncertainty and distance; however, the correlation coefficient is very close to zero and not statistically significant ($p > 0.05$).

The correlation analysis reveals that uncertainty can serve as a meaningful, although imperfect, indicator of test oracle reliability. For \gemini, the consistent and statistically significant correlations across tasks suggest that higher uncertainty is generally aligned with poorer oracle decisions, as reflected by larger distances from the ground truth. This indicates that, for this model, uncertainty estimates provide useful information about when its assessments should be treated with caution. In contrast, \openai exhibits weaker and less consistent correlation, with some tasks showing no statistically significant relationship between uncertainty and distance. This discrepancy suggests that the usefulness of uncertainty as a proxy for correctness might be model-dependent and influenced by how uncertainty is internally represented and exposed. Notably, the absence of significant correlations for certain tasks implies that low uncertainty does not always guarantee accurate predictions.

\begin{custombox}{Answer to RQ3}
Uncertainty and distance are slightly correlated, with \gemini showing consistent positive correlations and \openai exhibiting similar but even weaker trends.
\end{custombox}

\section{Threats to Validity}\label{sec:threats}

Threats to conclusion validity are mainly related to the stochastic nature of the models used, even when the temperature is set to 0. To mitigate these effects, we ran each task 10 times to account for variability and applied appropriate statistical tests in accordance with established guidelines in the literature~\cite{ArcuriICSE11}. 

Threats to internal validity arise from the models' parameter settings. We configured them identically to enable a fair comparison. For example, we set the temperature parameter to 0 for both models. Moreover, we used the same prompt template for both models.

Threats to construct validity arise from the evaluation metrics used. For task correctness and task quality assessment, we used well-established metrics from the machine learning domain. Moreover, to capture the extent to which predictive quality deviates from the ground truth, we defined a distance-based metric.

A threat to external validity could be that we use four tasks, around 1,000 videos, and two models. However, this is a sufficiently large number of videos extracted from three state-of-the-art vision language action model-based robots. Including a broader set of tasks and videos would strengthen our findings across diverse tasks, while evaluating additional VLM models would help assess how well the results generalize beyond those considered in this study.

Another external validity threat is that \approach has been tested only on simulation datasets. While \approach can be adapted to real-world use cases, how well it performs is an area of future investigation. It is expected that real-world settings will introduce additional challenges like sensor noise, motion blur, etc. These factors may further affect the results of task-correctness and task-quality. Therefore, our current claims are limited to simulation-recorded datasets. 

\section{Discussion} \label{sec:discussion}
This paper assessed the feasibility of using VLMs as test oracles for robotic tasks testing by analyzing task execution videos generated from simulation. Although we were unable to use open-source VLMs effectively, proprietary VLMs performed well in assessing task correctness and quality, with low uncertainty. Moreover, the results were consistent across multiple runs, indicating reliable performance. However, we observed that model uncertainty remained low even when the VLM failed at properly determining task correctness and quality, suggesting that the employed uncertainty metrics may not reliably reflect prediction confidence. As a result, we foresee the need to define new uncertainty metrics that can be reliably used to assess task correctness and quality. 

Moreover, we see potential to reduce costs using VLMs as a test oracle for robotic task testing. In our experiments, the per-test cost associated with monetary costs is approximately \$0.03 per assessment and inference time of approximately 7 seconds. Compared to manual assessment, where a human evaluates a task in real-time, and for the failing cases there is no need to finish the task assessment, \approach must wait for the simulation to complete, resulting in inference time and monetary cost. For a single video or a small set of videos, human assessment is generally faster and more reliable. However, as the number of videos to evaluate increases, the advantage of using \approach becomes more significant. Apart from being able to evaluate videos continuously for days, in environments where multiple robots are operating simultaneously, \approach can process several videos in parallel, limited only by the capacity of the VLM to handle multiple requests. Meanwhile, using \approach frees human evaluators to focus on other tasks during this time, further improving overall efficiency. In contrast, a human can evaluate only one video at a time, making \approach a more scalable solution for assessing thousands of tests. We note, however, that our current evaluation focused on videos generated from simulations. Consequently, the performance of \approach on videos from hardware-in-the-loop simulations or real-world robotic deployments remains untested and represents an important direction for future work. At this stage, our focus is on simulation-based testing, which is the primary setting for large-scale evaluation of robotic applications. Accordingly, \approach is designed to support automated test assessment at scale in simulation-based robotics CI/CD pipelines.

Finally, we anticipate that fine-tuning open-source VLMs on robotic task assessment data could improve their performance and reduce reliance on proprietary models, thereby reducing the monetary costs associated with proprietary models.

\section{Related work}\label{sec:related work}
Recent research have enabled robots to perform complex manipulation tasks by integrating multimodal perception with action generation. However, evaluating these systems remains challenging in open-world settings, where failures stem from perceptual ambiguities, environmental variability, or underspecified instructions. Most existing approaches rely on benchmark-based evaluations with symbolic oracles~\cite{james2020rlbench,li2024evaluating, liu2023libero, nasiriany2024robocasa}. For instance, VLATest~\cite{wang2025vlatest}, which automatically generates scenes to evaluate VLAs, and large-scale benchmarks like VLABench~\cite{zhang2025vlabench} and Nebula~\cite{peng2025nebula}, primarily report success rates focusing on final outcome correctness. However, these approaches offer limited insight into qualitative aspects such as motion quality, robustness, and perceptual grounding, while also relying on ad-hoc, task-specific test oracles. To address these limitations, Valle et al.~\cite{VLAvalleArXiv2025} proposed quality and uncertainty metrics for robotic task evaluation. In contrast to our approach, their method requires human intervention to set metric thresholds so that the metrics distinguish between different levels of execution quality.

A complementary research direction leverages Large Language Models (LLMs) and VLMs for failure detection and automated correction. REFLECT~\cite{liu2023reflect} converts multi-sensory robot actions into hierarchical textual summaries, enabling LLMs to explain failures and suggest corrective actions. RoboReflect~\cite{luo2025roboreflect} employs VLMs for reflective reasoning and trajectory planning adjustment in ambiguous grasping scenarios, while SC-VLA~\cite{li2024self} integrates fast action prediction with a slower VLM-based system for detecting and correcting errors via chain-of-thought reasoning. Code-as-Monitor~\cite{zhou2025code} leverages VLMs to generate code to monitor spatio-temporal constraints in both reactive and proactive modes of robotic tasks, and methods like RoboFAC~\cite{lu2025robofac} and AHA~\cite{duan2024aha} fine-tune VLMs for natural-language reasoning over failures and trajectory corrections. Other approaches~\cite{guo2024doremi, du2023vision} leverage VLMs for visual question answering, providing either binary task correctness or user-specific responses. While these methods demonstrate the utility of LLMs and VLMs for monitoring and recovery, they largely focus on failure correction or binary evaluation rather than assessment of execution quality.

In contrast to these approaches, \approach evaluates not only task success but also execution quality, and provides uncertainty estimation to support trust in its verdicts. Moreover, unlike approaches that use symbolic oracles, \approach can be easily adapted to other tasks, requiring no human intervention for threshold tuning or metric specification, but only for defining task quality requirements.

\section{Conclusion and Future Work}\label{sec:conclusion}
Test oracle assessment for testing robotic behavior to determine its correctness remains a manual and, therefore, time-consuming task. To address this challenge, this paper proposes a VLM–based approach for automated test oracle assessment that determines not only whether a robot's behavior is correct but also its quality based on recorded robot behavioral videos. Moreover, \approach explicitly quantifies uncertainty when assessing both task correctness and task quality, thereby providing a measure of how much the VLM's assessment can be trusted. We evaluate \approach using two VLMs, \openai and \gemini, across four robotic tasks, analyzing their ability to assess correctness and quality while accounting for uncertainty. 

As future work, we plan to evaluate World models like Cosmos from Nvidia~\cite{cosmosNvidiaArxiv}. These emerging models can support complex video analytics over large volumes of recorded and live video, enabling richer, more contextual understanding of visual content. Moreover, we will expand our evaluation to more VLMs and develop a voting-based mechanism to consolidate assessments from multiple VLMs.

\begin{acks}
This work is supported by the InnoGuard Marie Skłodowska-Curie Doctoral Network (Grant Agreement No. 101169233). P. Arcaini is supported by the ASPIRE grant No. JPMJAP2301, JST.
Pablo Valle and Aitor Arrieta are part of the Software and Systems Engineering research group of Mondragon Unibertsitatea (IT1519-22), supported by the Department of Education, Universities and Research of the Basque Country. Pablo Valle is supported by the Pre-doctoral Program for the Formation of Non-Doctoral Research Staff of the Education Department of the Basque Government (Grant n. PRE\_2025\_2\_0252). We used generative AI to generate some code for the manuscript (e.g., LaTeX tables), implementation, and experiments.
\end{acks}

\bibliographystyle{ACM-Reference-Format}
\bibliography{biblio}

\end{document}

%% file: images/prompt_binary.tex
\centering
\scriptsize
\begin{promptframe}
\begin{promptband}{ctxblue}
These images are ordered frames from a robot grasping video.\\
You are a precise evaluator of quality of robotic task performance from entire sequence.
\end{promptband}
\begin{promptband}{instorange}
Decide if the \textbf{\texttt{\{TASK\_INSTRUCTION\}}} performed by the robotic gripper arm was a Success or Failure in the given video.
\end{promptband}
\begin{promptband}{reqpink}
Output format: Return only valid JSON. No extra text. Use this schema exactly.
\begin{lstlisting}
{
  "status": "Successful" | "Failure"
}
\end{lstlisting}
\end{promptband}
\end{promptframe}

%% file: images/prompt_multi.tex
\centering
\scriptsize
\begin{promptframe}
\begin{promptband}{ctxblue}
These images are ordered frames from a robot grasping video.\\
Considering the entire sequence, you are a precise evaluator of the quality of robotic task performance from the entire sequence.
\end{promptband}

\begin{promptband}{inpgreen}
TASK: \textbf{\texttt{\{TASK\_INSTRUCTION\}}}
\end{promptband}

\begin{promptband}{instorange}
INSTRUCTION: Examine the GRIPPER and examine how it completes the Task.\\
Decide the overall quality of the task performed using only the decision rules below:
\end{promptband}

\begin{promptband}{rulespurple}
DECISION RULES in JSON format:
\begin{lstlisting}
{
  "High": "...",
  "Medium": "...",
  "Low": "..."
}
\end{lstlisting}
\end{promptband}

\begin{promptband}{reqpink}
Output format:\\
Return only valid json. No extra text. Use this schema exactly.
\begin{lstlisting}
{
  "quality": "high" | "medium" | "low"
}
\end{lstlisting}
\end{promptband}
\end{promptframe}

%% file: tables/table_rq1a_recall_precision_f1.tex
\resizebox{1\columnwidth}{!}{
\begin{tabular}{l l r rr rr r}
\toprule
\multirow{2}{*}{\textbf{VLM}} & \multirow{2}{*}{\textbf{Task}} &
\multicolumn{2}{c}{\textbf{Precision}} &
\multicolumn{2}{c}{\textbf{Recall}} &
\multicolumn{2}{c}{\textbf{F1}} \\
\cmidrule{3-8}
& &
\multicolumn{1}{c}{$m$} & \multicolumn{1}{c}{$\sigma$} &
\multicolumn{1}{c}{$m$} & \multicolumn{1}{c}{$\sigma$} &
\multicolumn{1}{c}{$m$} & \multicolumn{1}{c}{$\sigma$} \\
\midrule
\multirow{4}{*}{\textbf{\openai}} & \movenear
& 0.844 & 0.044
& 0.251 & 0.016
& 0.386 & 0.019 \\
& \pickup
& 0.775 & 0.007
& 0.541 & 0.018
& 0.637 & 0.013 \\
& \putin
& 0.980 & 0.000
& 0.739 & 0.027
& 0.842 & 0.018 \\
& \puton
& 0.959 & 0.015
& 0.746 & 0.018
& 0.839 & 0.013 \\
\midrule
\multirow{4}{*}{\textbf{\gemini}} & \movenear
&  0.561 & 0.014
&  0.730 & 0.000
&  0.635 & 0.009 \\
& \pickup
& 0.609 & 0.007
& 0.836 & 0.003
& 0.704 & 0.004 \\
& \putin
& 0.782 & 0.053
& 0.651 & 0.031
& 0.711 & 0.041 \\
& \puton
& 0.674 & 0.030
& 0.869 & 0.005
& 0.759 & 0.021 \\
\bottomrule
\end{tabular}
}

%% file: tables/table_rq1a_stats.tex
\resizebox{1\columnwidth}{!}{
\begin{tabular}{l r r r r r r}
\toprule
\multirow{2}{*}{\textbf{Task}}& 
\multicolumn{2}{c}{\textbf{Precision}} &
\multicolumn{2}{c}{\textbf{Recall}} &
\multicolumn{2}{c}{\textbf{F1}} \\
\cmidrule{2-7}
&
\multicolumn{1}{c}{$p$-value} & \multicolumn{1}{c}{\Atwelve} &
\multicolumn{1}{c}{$p$-value} & \multicolumn{1}{c}{\Atwelve} &
\multicolumn{1}{c}{$p$-value} & \multicolumn{1}{c}{\Atwelve} \\
\midrule
\texttt{\movenear} &
\multicolumn{1}{c}{0.000144} & \multicolumn{1}{c}{1.00} &
\multicolumn{1}{c}{$<0.0001$} & \multicolumn{1}{c}{0.00} &
\multicolumn{1}{c}{0.000144} & \multicolumn{1}{c}{0.00} \\
\texttt{\pickup} &
\multicolumn{1}{c}{0.000144} & \multicolumn{1}{c}{1.00} &
\multicolumn{1}{c}{0.000142} & \multicolumn{1}{c}{0.00} &
\multicolumn{1}{c}{0.000144} & \multicolumn{1}{c}{0.00} \\
\texttt{\putin} &
\multicolumn{1}{c}{0.000141} & \multicolumn{1}{c}{1.00} &
\multicolumn{1}{c}{0.000142} & \multicolumn{1}{c}{1.00} &
\multicolumn{1}{c}{0.000142} & \multicolumn{1}{c}{1.00} \\
\texttt{\puton} &
\multicolumn{1}{c}{0.000145} & \multicolumn{1}{c}{1.00} &
\multicolumn{1}{c}{0.000141} & \multicolumn{1}{c}{0.00} &
\multicolumn{1}{c}{0.000145} & \multicolumn{1}{c}{1.00} \\
\bottomrule
\end{tabular}
}

%% file: tables/table_rq1b_recall_precision_f1.tex
\resizebox{1\textwidth}{!}{
\begin{tabular}{l l c c c c c c c c c c c c c c}
\toprule
\multirow{2}{*}{\textbf{VLM}} & \multirow{2}{*}{\textbf{Task}} &
\multicolumn{2}{c}{\textbf{Precision-micro}} &
\multicolumn{2}{c}{\textbf{Precision-macro}} &
\multicolumn{2}{c}{\textbf{Recall-micro}} &
\multicolumn{2}{c}{\textbf{Recall-macro}} &
\multicolumn{2}{c}{\textbf{F1-micro}} &
\multicolumn{2}{c}{\textbf{F1-macro}} &
\multicolumn{2}{c}{\textbf{Distance}} \\
\cmidrule{3-16}
& &
$m$ & \multicolumn{1}{c}{$\sigma$} &
$m$ & \multicolumn{1}{c}{$\sigma$} &
$m$ & \multicolumn{1}{c}{$\sigma$} &
$m$ & \multicolumn{1}{c}{$\sigma$} &
$m$ & \multicolumn{1}{c}{$\sigma$} &
$m$ & \multicolumn{1}{c}{$\sigma$} &
$m$ & $\sigma$ \\
\midrule
\multirow{4}{*}{\textbf{\openai}} & \movenear
& 0.193 & \multicolumn{1}{c}{0.0232}
& 0.168 & \multicolumn{1}{c}{0.0365}
& 0.193 & \multicolumn{1}{c}{0.0232}
& 0.220 & \multicolumn{1}{c}{0.0208}
& 0.193 & \multicolumn{1}{c}{0.0232}
& 0.164 & \multicolumn{1}{c}{0.0264}
& 1.856 & 0.0482 \\
& \pickup
& 0.390 & \multicolumn{1}{c}{0.0175}
& 0.267 & \multicolumn{1}{c}{0.0273}
& 0.390 & \multicolumn{1}{c}{0.0175}
& 0.344 & \multicolumn{1}{c}{0.0175}
& 0.390 & \multicolumn{1}{c}{0.0175}
& 0.265 & \multicolumn{1}{c}{0.0191}
& 1.534 & 0.0389 \\
& \putin
& 0.386 & \multicolumn{1}{c}{0.0219}
& 0.266 & \multicolumn{1}{c}{0.0218}
& 0.386 & \multicolumn{1}{c}{0.0219}
& 0.301 & \multicolumn{1}{c}{0.0163}
& 0.386 & \multicolumn{1}{c}{0.0219}
& 0.270 & \multicolumn{1}{c}{0.0172}
& 0.847 & 0.0475 \\
& \puton
& 0.483 & \multicolumn{1}{c}{0.0214}
& 0.340 & \multicolumn{1}{c}{0.0379}
& 0.483 & \multicolumn{1}{c}{0.0214}
& 0.383 & \multicolumn{1}{c}{0.0179}
& 0.483 & \multicolumn{1}{c}{0.0214}
& 0.333 & \multicolumn{1}{c}{0.0231}
& 0.843 & 0.0345 \\
\midrule
\multirow{4}{*}{\textbf{\gemini}} & \movenear
& 0.205 & \multicolumn{1}{c}{0.0015}
& 0.152 & \multicolumn{1}{c}{0.0032}
& 0.205 & \multicolumn{1}{c}{0.0015}
& 0.262 & \multicolumn{1}{c}{0.0021}
& 0.205 & \multicolumn{1}{c}{0.0015}
& 0.155 & \multicolumn{1}{c}{0.0022}
& 2.506 & 0.0015 \\
& \pickup
& 0.222 & \multicolumn{1}{c}{0.0000}
& 0.141 & \multicolumn{1}{c}{0.0000}
& 0.222 & \multicolumn{1}{c}{0.0000}
& 0.270 & \multicolumn{1}{c}{0.0000}
& 0.222 & \multicolumn{1}{c}{0.0000}
& 0.150 & \multicolumn{1}{c}{0.0000}
& 2.418 & 0.0000 \\
& \putin
& 0.301 & \multicolumn{1}{c}{0.0000}
& 0.221 & \multicolumn{1}{c}{0.0000}
& 0.301 & \multicolumn{1}{c}{0.0000}
& 0.297 & \multicolumn{1}{c}{0.0000}
& 0.301 & \multicolumn{1}{c}{0.0000}
& 0.182 & \multicolumn{1}{c}{0.0000}
& 1.794 & 0.0000 \\
& \puton
& 0.276 & \multicolumn{1}{c}{0.0000}
& 0.217 & \multicolumn{1}{c}{0.0000}
& 0.276 & \multicolumn{1}{c}{0.0000}
& 0.322 & \multicolumn{1}{c}{0.0000}
& 0.276 & \multicolumn{1}{c}{0.0000}
& 0.195 & \multicolumn{1}{c}{0.0000}
& 2.153 & 0.0000 \\
\bottomrule
\end{tabular}}

%% file: tables/table_rq1b_stats.tex
\resizebox{1\textwidth}{!}{%
\begin{tabular}{l rrrrrrrrrrrrrr}
\toprule
\multirow{2}{*}{\textbf{Task}}&
\multicolumn{2}{c}{\textbf{Precision-micro}} &
\multicolumn{2}{c}{\textbf{Precision-macro}} &
\multicolumn{2}{c}{\textbf{Recall-micro}} &
\multicolumn{2}{c}{\textbf{Recall-macro}} &
\multicolumn{2}{c}{\textbf{F1-micro}} &
\multicolumn{2}{c}{\textbf{F1-macro}} &
\multicolumn{2}{c}{\textbf{Distance}}
\\\cmidrule{2-15}
&
$p$-value & \multicolumn{1}{c}{\Atwelve} &
$p$-value & \multicolumn{1}{c}{\Atwelve} &
$p$-value & \multicolumn{1}{c}{\Atwelve} &
$p$-value & \multicolumn{1}{c}{\Atwelve} &
$p$-value & \multicolumn{1}{c}{\Atwelve} &
$p$-value & \multicolumn{1}{c}{\Atwelve} &
$p$-value & \Atwelve \\
\midrule
\texttt{\movenear} 
& 0.0189 & \multicolumn{1}{c}{0.20}
& 0.1426 & \multicolumn{1}{c}{0.69}
& 0.0189 & \multicolumn{1}{c}{0.20}
& $<0.0001$ & \multicolumn{1}{c}{0.00}
& 0.0189 & \multicolumn{1}{c}{0.20}
& 0.0433 & \multicolumn{1}{c}{0.76}
& $<0.0001$ & 0.00 \\
\texttt{\pickup} 
& $<0.0001$ & \multicolumn{1}{c}{1.00}
& $<0.0001$ & \multicolumn{1}{c}{1.00}
& $<0.0001$ & \multicolumn{1}{c}{1.00}
& $<0.0001$ & \multicolumn{1}{c}{1.00}
& $<0.0001$ & \multicolumn{1}{c}{1.00}
& $<0.0001$ & \multicolumn{1}{c}{1.00}
& $<0.0001$ & 0.00 \\
\texttt{\putin} 
& $<0.0001$ & \multicolumn{1}{c}{1.00}
& $<0.0001$ & \multicolumn{1}{c}{1.00}
& $<0.0001$ & \multicolumn{1}{c}{1.00}
& 1.0000 & \multicolumn{1}{c}{0.50}
& $<0.0001$ & \multicolumn{1}{c}{1.00}
& $<0.0001$ & \multicolumn{1}{c}{1.00}
& $<0.0001$ & 0.00 \\

\texttt{\puton}
& $<0.0001$ & \multicolumn{1}{c}{1.00}
& $<0.0001$ & \multicolumn{1}{c}{1.00}
& $<0.0001$ & \multicolumn{1}{c}{1.00}
& $<0.0001$ & \multicolumn{1}{c}{1.00}
& $<0.0001$ & \multicolumn{1}{c}{1.00}
& $<0.0001$ & \multicolumn{1}{c}{1.00}
& $<0.0001$ & 0.00 \\
\bottomrule
\end{tabular}}

%% file: tables/table_rq2_binary_uncertainty.tex
\resizebox{1\columnwidth}{!}{
\begin{tabular}{l l r rr rr r}
\toprule
\multirow{2}{*}{\textbf{VLM}} & \multirow{2}{*}{\textbf{Task}} &
\multicolumn{2}{c}{\textbf{DeepGini} $\downarrow$} &
\multicolumn{2}{c}{\textbf{Entropy} $\downarrow$ } &
\multicolumn{2}{c}{\textbf{MSP} $\uparrow$} \\
\cmidrule{3-8}
& &
\multicolumn{1}{c}{$m$} & \multicolumn{1}{c}{$\sigma$} &
\multicolumn{1}{c}{$m$} & \multicolumn{1}{c}{$\sigma$} &
\multicolumn{1}{c}{$m$} & \multicolumn{1}{c}{$\sigma$} \\
\midrule
\multirow{4}{*}{\textbf{\openai}} & \movenear & 0.0733 & 0.0089 & 0.0311 & 0.0041 & 0.9789 & 0.0027 \\
& \pickup & 0.0445 & 0.0050 & 0.0192 & 0.0023 & 0.9865 & 0.0017 \\
& \putin & 0.0222 & 0.0064 & 0.0090 & 0.0029 & 0.9942 & 0.0021 \\
& \puton & 0.0447 & 0.0071 & 0.0193 & 0.0036 & 0.9865 & 0.0031 \\
\midrule
\multirow{4}{*}{\textbf{\gemini}} & \movenear & 0.0007 & 0.0001 & 0.0001 & 0.0000 & 0.9999 & 0.0000 \\
& \pickup & 0.0003 & 0.0001 & 0.0001 & 0.0000 & 1.0000 & 0.0000 \\
& \putin & 0.0006 & 0.0001 & 0.0001 & 0.0000 & 1.0000 & 0.0000 \\
& \puton & 0.0004 & 0.0001 & 0.0001 & 0.0000 & 1.0000 & 0.0000 \\
\bottomrule
\end{tabular}
}

%% file: tables/table_rq2_multiclass_uncertainty.tex
\resizebox{1\columnwidth}{!}{
\begin{tabular}{l l r rr rr r}
\toprule
\multirow{2}{*}{\textbf{VLM}} & \multirow{2}{*}{\textbf{Task}} &
\multicolumn{2}{c}{\textbf{DeepGini} $\downarrow$} &
\multicolumn{2}{c}{\textbf{Entropy} $\downarrow$} &
\multicolumn{2}{c}{\textbf{MSP} $\uparrow$} \\
\cmidrule{3-8}
& &
\multicolumn{1}{c}{$m$} & \multicolumn{1}{c}{$\sigma$} &
\multicolumn{1}{c}{$m$} & \multicolumn{1}{c}{$\sigma$} &
\multicolumn{1}{c}{$m$} & \multicolumn{1}{c}{$\sigma$} \\
\midrule
\multirow{4}{*}{\textbf{\openai}} & \movenear & 0.0978 & 0.0369 & 0.1599 & 0.0568 & 0.9281 & 0.0303 \\
& \pickup  & 0.0322 & 0.0174 & 0.0536 & 0.0274 & 0.9770 & 0.0135 \\
& \putin   & 0.0468 & 0.0315 & 0.0812 & 0.0502 & 0.9683 & 0.0242 \\
& \puton   & 0.0585 & 0.0176 & 0.0974 & 0.0246 & 0.9592 & 0.0152 \\
\midrule
\multirow{4}{*}{\textbf{\gemini}} & \movenear & 0.0008 & 0.0005 & 0.0026 & 0.0014 & 0.9996 & 0.0003 \\
& \pickup  & 0.0013 & 0.0015 & 0.0032 & 0.0026 & 0.9992 & 0.0009 \\
& \putin   & 0.0012 & 0.0011 & 0.0033 & 0.0023 & 0.9994 & 0.0006 \\
& \puton   & 0.0002 & 0.0001 & 0.0010 & 0.0004 & 0.9999 & 0.0000 \\
\bottomrule
\end{tabular}
}

%% file: tables/table_rq3_spearman.tex
\resizebox{1\columnwidth}{!}{
\begin{tabular}{l l r rr rr r}
\toprule
\multirow{2}{*}{\textbf{VLM}} & \multirow{2}{*}{\textbf{Task}} &
\multicolumn{2}{c}{\textbf{DeepGini} } &
\multicolumn{2}{c}{\textbf{Entropy} } &
\multicolumn{2}{c}{\textbf{MSP} } \\
\cmidrule{3-8}
& &
\multicolumn{1}{c}{\textit{$p$-value}} & \multicolumn{1}{c}{\textit{$\rho$}} &
\multicolumn{1}{c}{\textit{$p$-value}} & \multicolumn{1}{c}{\textit{$\rho$}} &
\multicolumn{1}{c}{\textit{$p$-value}} & \multicolumn{1}{c}{\textit{$\rho$}}\\
\midrule
\multirow{4}{*}{\textbf{\openai}} & \movenear
& 0.8105 & 0.0088
& 0.8699 & 0.0060
& 0.7884 & $-0.0099$ \\
& \pickup
& $<0.0001$ & 0.2193
& $<0.0001$ & 0.2193
& $<0.0001$ & $-0.2194$ \\
& \putin
& 0.6558 & $-0.0187$
& 0.6569 & $-0.0186$
& 0.6573 & 0.0186 \\
& \puton
& $<0.0001$ & 0.1657
& $<0.0001$ & 0.1666
& $<0.0001$ & $-0.1657$ \\
\midrule
\multirow{4}{*}{\textbf{\gemini}} & \movenear
& 0.0003 & 0.0776
& 0.0003 & 0.0784
& 0.0003 & $-0.0776$ \\
& \pickup
& 0.0077 & 0.0482
& 0.0055 & 0.0502
& 0.0077 & $-0.0482$ \\
& \putin
& $<0.0001$ & 0.3120
& $<0.0001$ & 0.3120
& $<0.0001$ & $-0.3120$ \\
& \puton
& 0.0297 & 0.0539
& 0.0192 & 0.0580
& 0.0297 & $-0.0539$ \\
\bottomrule
\end{tabular}
}

%% file: biblio.bib
@article{gpt4Arxiv,
author = {OpenAI},
title = {{GPT-4} Technical Report},
journal = {CoRR},
volume = {abs/2303.08774},
year = {2023},
url = {https://doi.org/10.48550/arXiv.2303.08774},
doi = {10.48550/ARXIV.2303.08774},
eprinttype = {arXiv},
eprint = {2303.08774}
}

@article{qwenArxiv,
author = {An Yang and Anfeng Li and Baosong Yang and Beichen Zhang and Binyuan Hui and Bo Zheng and Bowen Yu and Chang Gao and Chengen Huang and Chenxu Lv and Chujie Zheng and Dayiheng Liu and Fan Zhou and Fei Huang and Feng Hu and Hao Ge and Haoran Wei and Huan Lin and Jialong Tang and Jian Yang and Jianhong Tu and Jianwei Zhang and Jian Yang and Jiaxi Yang and Jingren Zhou and Junyang Lin and Kai Dang and Keqin Bao and Kexin Yang and Le Yu and Lianghao Deng and Mei Li and Mingfeng Xue and Mingze Li and Pei Zhang and Peng Wang and Qin Zhu and Rui Men and Ruize Gao and Shixuan Liu and Shuang Luo and Tianhao Li and Tianyi Tang and Wenbiao Yin and Xingzhang Ren and Xinyu Wang and Xinyu Zhang and Xuancheng Ren and Yang Fan and Yang Su and Yichang Zhang and Yinger Zhang and Yu Wan and Yuqiong Liu and Zekun Wang and Zeyu Cui and Zhenru Zhang and Zhipeng Zhou and Zihan Qiu},
title = {Qwen3 Technical Report},
journal = {CoRR},
volume = {abs/2505.09388},
year = {2025},
url = {https://doi.org/10.48550/arXiv.2505.09388},
doi = {10.48550/ARXIV.2505.09388},
eprinttype = {arXiv},
eprint = {2505.09388}
}

@article{SmolVLMArxiv,
author = {Andr{\'{e}}s Marafioti and Orr Zohar and Miquel Farr{\'{e}} and Merve Noyan and Elie Bakouch and Pedro Cuenca and Cyril Zakka and Loubna Ben Allal and Anton Lozhkov and Nouamane Tazi and Vaibhav Srivastav and Joshua Lochner and Hugo Larcher and Mathieu Morlon and Lewis Tunstall and Leandro {von Werra} and Thomas Wolf},
title = {{SmolVLM}: Redefining small and efficient multimodal models},
journal = {CoRR},
volume = {abs/2504.05299},
year = {2025},
url = {https://doi.org/10.48550/arXiv.2504.05299},
doi = {10.48550/ARXIV.2504.05299},
eprinttype = {arXiv},
eprint = {2504.05299}
}

@inproceedings{EagleVLM,
author = {Min Shi and Fuxiao Liu and Shihao Wang and Shijia Liao and Subhashree Radhakrishnan and Yilin Zhao and De{-}An Huang and Hongxu Yin and Karan Sapra and Yaser Yacoob and Humphrey Shi and Bryan Catanzaro and Andrew Tao and Jan Kautz and Zhiding Yu and Guilin Liu},
title = {Eagle: {Exploring} The Design Space for Multimodal LLMs with Mixture of Encoders},
booktitle = {The Thirteenth International Conference on Learning Representations, {ICLR} 2025, Singapore, April 24-28, 2025},
publisher = {OpenReview.net},
year = {2025},
url = {https://openreview.net/forum?id=Y2RW9EVwhT}
}

@article{LLaVANextarxiv,
author = {Feng Li and Renrui Zhang and Hao Zhang and Yuanhan Zhang and Bo Li and Wei Li and Zejun Ma and Chunyuan Li},
title = {{LLaVA-NeXT-Interleave}: Tackling Multi-image, Video, and 3D in Large Multimodal Models},
journal = {CoRR},
volume = {abs/2407.07895},
year = {2024},
url = {https://doi.org/10.48550/arXiv.2407.07895},
doi = {10.48550/ARXIV.2407.07895},
eprinttype = {arXiv},
eprint = {2407.07895}
}

@article{comanici2025gemini,
author = {{Gemini Team}},
title = {Gemini 2.5: Pushing the Frontier with Advanced Reasoning, Multimodality, Long Context, and Next Generation Agentic Capabilities},
journal = {CoRR},
volume = {abs/2507.06261},
year = {2025},
url = {https://doi.org/10.48550/arXiv.2507.06261},
doi = {10.48550/ARXIV.2507.06261},
eprinttype = {arXiv},
eprint = {2507.06261}
}

@article{VLAvalleArXiv2025,
author = {Valle, Pablo and Lu, Chengjie and Ali, Shaukat and Arrieta, Aitor},
title = {Evaluating Uncertainty and Quality of Visual Language Action-enabled Robots},
journal = {CoRR},
volume = {abs/2507.17049},
year = {2025},
url = {https://doi.org/10.48550/arXiv.2507.17049},
doi = {10.48550/ARXIV.2507.17049},
eprinttype = {arXiv},
eprint = {2507.17049}
}

@inproceedings{radford2021learning,
title = {Learning Transferable Visual Models From Natural Language Supervision},
author = {Radford, Alec and Kim, Jong Wook and Hallacy, Chris and Ramesh, Aditya and Goh, Gabriel and Agarwal, Sandhini and Sastry, Girish and Askell, Amanda and Mishkin, Pamela and Clark, Jack and Krueger, Gretchen and Sutskever, Ilya},
booktitle = {Proceedings of the 38th International Conference on Machine Learning},
pages = {8748--8763},
year = {2021},
editor = {Meila, Marina and Zhang, Tong},
volume = {139},
series = {Proceedings of Machine Learning Research},
month = {18--24 Jul},
publisher = {PMLR},
pdf_ = {http://proceedings.mlr.press/v139/radford21a/radford21a.pdf},
url = {https://proceedings.mlr.press/v139/radford21a.html}
}

@article{wang2022git,
author = {Jianfeng Wang and Zhengyuan Yang and Xiaowei Hu and Linjie Li and Kevin Lin and Zhe Gan and Zicheng Liu and Ce Liu and Lijuan Wang},
title = {{GIT:} {A} Generative Image-to-text Transformer for Vision and Language},
journal = {CoRR},
volume = {abs/2205.14100},
year = {2022},
url = {https://doi.org/10.48550/arXiv.2205.14100},
doi = {10.48550/ARXIV.2205.14100},
eprinttype = {arXiv},
eprint = {2205.14100}
}

@inproceedings{liu2023visual,
author = {Liu, Haotian and Li, Chunyuan and Wu, Qingyang and Lee, Yong Jae},
title = {Visual instruction tuning},
year = {2023},
publisher = {Curran Associates Inc.},
address = {Red Hook, NY, USA},
booktitle = {Proceedings of the 37th International Conference on Neural Information Processing Systems},
articleno = {1516},
numpages = {25},
location = {New Orleans, LA, USA},
series = {NIPS '23}
}

@inproceedings{zitkovich2023rt,
title = {{RT-2}: Vision-Language-Action Models Transfer Web Knowledge to Robotic Control},
author = {Zitkovich, Brianna and Yu, Tianhe and Xu, Sichun and Xu, Peng and Xiao, Ted and Xia, Fei and Wu, Jialin and Wohlhart, Paul and Welker, Stefan and Wahid, Ayzaan and Vuong, Quan and Vanhoucke, Vincent and Tran, Huong and Soricut, Radu and Singh, Anikait and Singh, Jaspiar and Sermanet, Pierre and Sanketi, Pannag R. and Salazar, Grecia and Ryoo, Michael S. and Reymann, Krista and Rao, Kanishka and Pertsch, Karl and Mordatch, Igor and Michalewski, Henryk and Lu, Yao and Levine, Sergey and Lee, Lisa and Lee, Tsang-Wei Edward and Leal, Isabel and Kuang, Yuheng and Kalashnikov, Dmitry and Julian, Ryan and Joshi, Nikhil J. and Irpan, Alex and Ichter, Brian and Hsu, Jasmine and Herzog, Alexander and Hausman, Karol and Gopalakrishnan, Keerthana and Fu, Chuyuan and Florence, Pete and Finn, Chelsea and Dubey, Kumar Avinava and Driess, Danny and Ding, Tianli and Choromanski, Krzysztof Marcin and Chen, Xi and Chebotar, Yevgen and Carbajal, Justice and Brown, Noah and Brohan, Anthony and Arenas, Montserrat Gonzalez and Han, Kehang},
booktitle = {Proceedings of The 7th Conference on Robot Learning},
pages = {2165--2183},
year = {2023},
editor = {Tan, Jie and Toussaint, Marc and Darvish, Kourosh},
volume = {229},
series = {Proceedings of Machine Learning Research},
month = {06--09 Nov},
publisher = {PMLR},
pdf_ = {https://proceedings.mlr.press/v229/zitkovich23a/zitkovich23a.pdf},
url = {https://proceedings.mlr.press/v229/zitkovich23a.html}
}

@article{nvidia2025gr00tn1openfoundation,
author = {Johan Bjorck and Fernando Casta{\~{n}}eda and Nikita Cherniadev and Xingye Da and Runyu Ding and Linxi Fan and Yu Fang and Dieter Fox and Fengyuan Hu and Spencer Huang and Joel Jang and Zhenyu Jiang and Jan Kautz and Kaushil Kundalia and Lawrence Lao and Zhiqi Li and Zongyu Lin and Kevin Lin and Guilin Liu and Edith LLontop and Loic Magne and Ajay Mandlekar and Avnish Narayan and Soroush Nasiriany and Scott Reed and You Liang Tan and Guanzhi Wang and Zu Wang and Jing Wang and Qi Wang and Jiannan Xiang and Yuqi Xie and Yinzhen Xu and Zhenjia Xu and Seonghyeon Ye and Zhiding Yu and Ao Zhang and Hao Zhang and Yizhou Zhao and Ruijie Zheng and Yuke Zhu},
title = {{GR00T} {N1:} An Open Foundation Model for Generalist Humanoid Robots},
journal = {CoRR},
volume = {abs/2503.14734},
year = {2025},
url = {https://doi.org/10.48550/arXiv.2503.14734},
doi = {10.48550/ARXIV.2503.14734},
eprinttype = {arXiv},
eprint = {2503.14734}
}

@article{mangiafico2016summary,
title = {Summary and analysis of extension program evaluation in {R}},
author = {Mangiafico, Salvatore S},
journal = {Rutgers Cooperative Extension: New Brunswick, NJ, USA},
volume = {125},
pages = {16--22},
year = {2016}
}

@inproceedings{li2024evaluating,
title = {Evaluating Real-World Robot Manipulation Policies in Simulation},
author = {Li, Xuanlin and Hsu, Kyle and Gu, Jiayuan and Mees, Oier and Pertsch, Karl and Walke, Homer Rich and Fu, Chuyuan and Lunawat, Ishikaa and Sieh, Isabel and Kirmani, Sean and Levine, Sergey and Wu, Jiajun and Finn, Chelsea and Su, Hao and Vuong, Quan and Xiao, Ted},
booktitle = {Proceedings of The 8th Conference on Robot Learning},
pages = {3705--3728},
year = {2025},
editor = {Agrawal, Pulkit and Kroemer, Oliver and Burgard, Wolfram},
volume = {270},
series = {Proceedings of Machine Learning Research},
month = {06--09 Nov},
publisher = {PMLR},
pdf_ = {https://raw.githubusercontent.com/mlresearch/v270/main/assets/li25c/li25c.pdf},
url = {https://proceedings.mlr.press/v270/li25c.html}
}

@inproceedings{nasiriany2024robocasa,
author = {Soroush Nasiriany and Abhiram Maddukuri and Lance Zhang and Adeet Parikh and Aaron Lo and Abhishek Joshi and Ajay Mandlekar and Yuke Zhu},
editor = {Dana Kulic and Gentiane Venture and {Kostas E.} Bekris and Enrique Coronado},
title = {{RoboCasa}: Large-Scale Simulation of Household Tasks for GeneralistRobots},
booktitle = {Robotics: Science and Systems XX, Delft, The Netherlands, July 15-19, 2024},
year = {2024},
url = {https://doi.org/10.15607/RSS.2024.XX.050},
doi = {10.15607/RSS.2024.XX.050}
}

@inproceedings{liu2023libero,
author = {Liu, Bo and Zhu, Yifeng and Gao, Chongkai and Feng, Yihao and Liu, Qiang and Zhu, Yuke and Stone, Peter},
title = {{LIBERO}: benchmarking knowledge transfer for lifelong robot learning},
year = {2023},
publisher = {Curran Associates Inc.},
address = {Red Hook, NY, USA},
booktitle = {Proceedings of the 37th International Conference on Neural Information Processing Systems},
articleno = {1939},
numpages = {16},
location = {New Orleans, LA, USA},
series = {NIPS '23}
}

@inproceedings{liu2023reflect,
title = {{REFLECT}: Summarizing Robot Experiences for Failure Explanation and Correction},
author = {Liu, Zeyi and Bahety, Arpit and Song, Shuran},
booktitle = {Proceedings of The 7th Conference on Robot Learning},
pages = {3468--3484},
year = {2023},
editor = {Tan, Jie and Toussaint, Marc and Darvish, Kourosh},
volume = {229},
series = {Proceedings of Machine Learning Research},
month = {06--09 Nov},
publisher = {PMLR},
pdf_ = {https://proceedings.mlr.press/v229/liu23g/liu23g.pdf},
url = {https://proceedings.mlr.press/v229/liu23g.html}
}

@article{james2020rlbench,
author = {James, Stephen and Ma, Zicong and Arrojo, David Rovick and Davison, Andrew J.},
journal = {IEEE Robotics and Automation Letters},
title = {{RLBench}: The Robot Learning Benchmark \& Learning Environment},
year = {2020},
volume = {5},
number = {2},
pages = {3019--3026},
doi = {10.1109/LRA.2020.2974707}
}

@article{qu2025spatialvla,
author = {Delin Qu and Haoming Song and Qizhi Chen and Yuanqi Yao and Xinyi Ye and Yan Ding and Zhigang Wang and JiaYuan Gu and Bin Zhao and Dong Wang and Xuelong Li},
title = {{SpatialVLA}: Exploring Spatial Representations for Visual-Language-Action Model},
journal = {CoRR},
volume = {abs/2501.15830},
year = {2025},
url = {https://doi.org/10.48550/arXiv.2501.15830},
doi = {10.48550/ARXIV.2501.15830},
eprinttype = {arXiv},
eprint = {2501.15830}
}

@article{luo2025roboreflect,
title = {{RoboReflect}: A Robotic Reflective Reasoning Framework for Grasping Ambiguous-Condition Objects},
author = {Luo, Zhen and Yang, Yixuan and Zhang, Yanfu and Zheng, Feng},
journal = {CoRR},
volume = {abs/2501.09307},
year = {2025},
url = {https://doi.org/10.48550/arXiv.2501.09307},
doi = {10.48550/ARXIV.2501.09307},
eprinttype = {arXiv},
eprint = {2501.09307}
}

@article{li2024self,
title = {A Self-Correcting Vision-Language-Action Model for Fast and Slow System Manipulation},
author = {Chenxuan Li and Jiaming Liu and Guanqun Wang and Xiaoqi Li and Sixiang Chen and Liang Heng and Chuyan Xiong and Jiaxin Ge and Renrui Zhang and Kaichen Zhou and Shanghang Zhang},
journal = {CoRR},
volume = {abs/2405.17418},
year = {2025},
eprint = {2405.17418},
archivePrefix = {arXiv},
primaryClass = {cs.CV},
url = {https://doi.org/10.48550/arXiv.2405.17418},
doi = {10.48550/ARXIV.2405.17418}
}

@inproceedings{duan2024aha,
author = {Jiafei Duan and Wilbert Pumacay and Nishanth Kumar and {Yi Ru} Wang and Shulin Tian and Wentao Yuan and Ranjay Krishna and Dieter Fox and Ajay Mandlekar and Yijie Guo},
title = {{AHA:} {A} Vision-Language-Model for Detecting and Reasoning Over Failures in Robotic Manipulation},
booktitle = {The Thirteenth International Conference on Learning Representations, {ICLR} 2025, Singapore, April 24-28, 2025},
publisher = {OpenReview.net},
year = {2025},
url = {https://openreview.net/forum?id=JVkdSi7Ekg}
}

@article{lu2025robofac,
author = {Weifeng Lu and Minghao Ye and Zewei Ye and Ruihan Tao and Shuo Yang and Bo Zhao},
title = {{RoboFAC}: {A} Comprehensive Framework for Robotic Failure Analysis
and Correction},
journal = {CoRR},
volume = {abs/2505.12224},
year = {2025},
url = {https://doi.org/10.48550/arXiv.2505.12224},
doi = {10.48550/ARXIV.2505.12224},
eprinttype = {arXiv},
eprint = {2505.12224}
}

@inproceedings{zhou2025code,
title = {Code-as-monitor: Constraint-aware visual programming for reactive and proactive robotic failure detection},
author = {Zhou, Enshen and Su, Qi and Chi, Cheng and Zhang, Zhizheng and Wang, Zhongyuan and Huang, Tiejun and Sheng, Lu and Wang, He},
booktitle = {Proceedings of the Computer Vision and Pattern Recognition Conference},
pages = {6919--6929},
year = {2025}
}

@inproceedings{du2023vision,
title = {Vision-Language Models as Success Detectors},
author = {Du, Yuqing and Konyushkova, Ksenia and Denil, Misha and Raju, Akhil and Landon, Jessica and Hill, Felix and de Freitas, Nando and Cabi, Serkan},
booktitle = {Proceedings of The 2nd Conference on Lifelong Learning Agents},
pages = {120--136},
year = {2023},
editor = {Chandar, Sarath and Pascanu, Razvan and Sedghi, Hanie and Precup, Doina},
volume = {232},
series = {Proceedings of Machine Learning Research},
month = {22--25 Aug},
publisher = {PMLR},
pdf_ = {https://proceedings.mlr.press/v232/du23b/du23b.pdf},
url = {https://proceedings.mlr.press/v232/du23b.html}
}

@inproceedings{guo2024doremi,
author = {Guo, Yanjiang and Wang, Yen-Jen and Zha, Lihan and Chen, Jianyu},
booktitle = {2024 IEEE/RSJ International Conference on Intelligent Robots and Systems (IROS)},
title = {{DoReMi}: Grounding Language Model by Detecting and Recovering from Plan-Execution Misalignment},
year = {2024},
volume = {},
number = {},
pages = {12124--12131},
doi = {10.1109/IROS58592.2024.10802284}
}

@inproceedings{li2024robonurse,
author = {Li, Shunlei and Wang, Jin and Dai, Rui and Ma, Wanyu and Ng, Wing Yin and Hu, Yingbai and Li, Zheng},
booktitle = {2025 IEEE/RSJ International Conference on Intelligent Robots and Systems (IROS)},
title = {{RoboNurse-VLA}: {R}obotic Scrub Nurse System based on Vision-Language-Action Model},
year = {2025},
volume = {},
number = {},
pages = {3986--3993},
doi = {10.1109/IROS60139.2025.11246030}
}

@article{peng2025nebula,
author = {Jierui Peng and Yanyan Zhang and Yicheng Duan and Tuo Liang and Vipin Chaudhary and Yu Yin},
title = {{NEBULA:} {Do} We Evaluate Vision-Language-Action Agents Correctly?},
journal = {CoRR},
volume = {abs/2510.16263},
year = {2025},
url = {https://doi.org/10.48550/arXiv.2510.16263},
doi = {10.48550/ARXIV.2510.16263},
eprinttype = {arXiv},
eprint = {2510.16263}
}

@article{zhang2025vlabench,
author = {Shiduo Zhang and Zhe Xu and Peiju Liu and Xiaopeng Yu and Yuan Li and Qinghui Gao and Zhaoye Fei and Zhangyue Yin and Zuxuan Wu and Yu{-}Gang Jiang and Xipeng Qiu},
title = {{VLABench}: {A} Large-Scale Benchmark for Language-Conditioned Robotics Manipulation with Long-Horizon Reasoning Tasks},
journal = {CoRR},
volume = {abs/2412.18194},
year = {2024},
url = {https://doi.org/10.48550/arXiv.2412.18194},
doi = {10.48550/ARXIV.2412.18194},
eprinttype = {arXiv},
eprint = {2412.18194}
}

@article{pantalone2021robot,
title = {Robot-assisted surgery in space: pros and cons. {A} review from the surgeon’s point of view},
volume = {7},
issn = {2373-8065},
url = {https://doi.org/10.1038/s41526-021-00183-3},
doi = {10.1038/s41526-021-00183-3},
number = {1},
journal = {npj Microgravity},
author = {Pantalone, Desirè and Faini, Giulia Satu and Cialdai, Francesca and Sereni, Elettra and Bacci, Stefano and Bani, Daniele and Bernini, Marco and Pratesi, Carlo and Stefàno, PierLuigi and Orzalesi, Lorenzo and Balsamo, Michele and Zolesi, Valfredo and Monici, Monica},
month = dec,
year = {2021},
pages = {56}
}

@article{wang2025vlatest,
author = {Wang, Zhijie and Zhou, Zhehua and Song, Jiayang and Huang, Yuheng and Shu, Zhan and Ma, Lei},
title = {{VLATest}: Testing and Evaluating Vision-Language-Action Models for Robotic Manipulation},
year = {2025},
issue_date = {July 2025},
publisher = {Association for Computing Machinery},
address = {New York, NY, USA},
volume = {2},
number = {FSE},
url = {https://doi.org/10.1145/3729343},
doi = {10.1145/3729343},
journal = {Proc. ACM Softw. Eng.},
month = jun,
articleno = {FSE073},
numpages = {24}
}

@article{rodriguez2021human,
author = {Rodríguez-Guerra, Diego and Sorrosal, Gorka and Cabanes, Itziar and Calleja, Carlos},
journal = {IEEE Access},
title = {Human-Robot Interaction Review: Challenges and Solutions for Modern Industrial Environments},
year = {2021},
volume = {9},
number = {},
pages = {108557--108578},
doi = {10.1109/ACCESS.2021.3099287}
}

@article{asif2025rapid,
title = {Rapid and automated configuration of robot manufacturing cells},
journal = {Robotics and Computer-Integrated Manufacturing},
volume = {92},
pages = {102862},
year = {2025},
issn = {0736-5845},
doi = {10.1016/j.rcim.2024.102862},
url = {https://www.sciencedirect.com/science/article/pii/S0736584524001492},
author = {Seemal Asif and Mikel Bueno and Pedro Ferreira and Paul Anandan and Ze Zhang and Yue Yao and Gautham Ragunathan and Lloyd Tinkler and Masoud Sotoodeh-Bahraini and Niels Lohse and Phil Webb and Windo Hutabarat and Ashutosh Tiwari}
}

@article{monostori2016cyber,
author = {Monostori, L{\'a}szl{\'o} and K{\'a}d{\'a}r, Botond and Bauernhansl, Thomas and Kondoh, Shinsuke and Kumara, Soundar and Reinhart, Gunther and Sauer, Olaf and Schuh, Gunther and Sihn, Wilfried and Ueda, Kenichi},
title = {Cyber-physical systems in manufacturing},
journal = {CIRP Annals},
volume = {65},
number = {2},
pages = {621--641},
year = {2016},
issn = {0007-8506},
doi = {10.1016/j.cirp.2016.06.005},
url = {https://www.sciencedirect.com/science/article/pii/S0007850616301974}
}

@article{dey2018medical,
author = {Dey, Nilanjan and Ashour, Amira S. and Shi, Fuqian and Fong, Simon James and Tavares, Jo\~{a}o Manuel R. S.},
title = {Medical cyber-physical systems: A survey},
year = {2018},
issue_date = {Apr 2018},
publisher = {Plenum Press},
address = {USA},
volume = {42},
number = {4},
issn = {0148-5598},
url = {https://doi.org/10.1007/s10916-018-0921-x},
doi = {10.1007/s10916-018-0921-x},
journal = {J. Med. Syst.},
month = mar,
numpages = {13}
}

@inproceedings{kato2018autoware,
author = {Kato, Shinpei and Tokunaga, Shota and Maruyama, Yuya and Maeda, Seiya and Hirabayashi, Manato and Kitsukawa, Yuki and Monrroy, Abraham and Ando, Tomohito and Fujii, Yusuke and Azumi, Takuya},
title = {Autoware on board: enabling autonomous vehicles with embedded systems},
year = {2018},
isbn = {9781538653012},
publisher = {IEEE Press},
url = {https://doi.org/10.1109/ICCPS.2018.00035},
doi = {10.1109/ICCPS.2018.00035},
booktitle = {Proceedings of the 9th ACM/IEEE International Conference on Cyber-Physical Systems},
pages = {287--296},
numpages = {10},
location = {Porto, Portugal},
series = {ICCPS '18}
}

@article{derler2011modeling,
title = {Modeling cyber--physical systems},
author = {Derler, Patricia and Lee, Edward A and Vincentelli, Alberto Sangiovanni},
journal = {Proceedings of the IEEE},
volume = {100},
number = {1},
pages = {13--28},
year = {2011},
publisher = {IEEE}
}

@article{baheti2011cyber,
title = {Cyber-physical systems},
author = {Baheti, Radhakisan and Gill, Helen},
journal = {The impact of control technology},
volume = {12},
number = {1},
pages = {161--166},
year = {2011}
}

@article{black2024pi0visionlanguageactionflowmodel,
title = {$\pi_0$: A Vision-Language-Action Flow Model for General Robot Control},
author = {Kevin Black and Noah Brown and Danny Driess and Adnan Esmail and Michael Equi and Chelsea Finn and Niccolo Fusai and Lachy Groom and Karol Hausman and Brian Ichter and Szymon Jakubczak and Tim Jones and Liyiming Ke and Sergey Levine and Adrian Li-Bell and Mohith Mothukuri and Suraj Nair and Karl Pertsch and Lucy Xiaoyang Shi and James Tanner and Quan Vuong and Anna Walling and Haohuan Wang and Ury Zhilinsky},
journal = {CoRR},
volume = {abs/2410.24164},
year = {2024},
url = {https://doi.org/10.48550/arXiv.2410.24164},
doi = {10.48550/ARXIV.2410.24164},
eprinttype = {arXiv},
eprint = {2410.24164}
}

@inproceedings{kim2024openvla,
title = {{OpenVLA}: An Open-Source Vision-Language-Action Model},
author = {Kim, Moo Jin and Pertsch, Karl and Karamcheti, Siddharth and Xiao, Ted and Balakrishna, Ashwin and Nair, Suraj and Rafailov, Rafael and Foster, Ethan P and Sanketi, Pannag R and Vuong, Quan and Kollar, Thomas and Burchfiel, Benjamin and Tedrake, Russ and Sadigh, Dorsa and Levine, Sergey and Liang, Percy and Finn, Chelsea},
booktitle = {Proceedings of The 8th Conference on Robot Learning},
pages = {2679--2713},
year = {2025},
editor = {Agrawal, Pulkit and Kroemer, Oliver and Burgard, Wolfram},
volume = {270},
series = {Proceedings of Machine Learning Research},
month = {06--09 Nov},
publisher = {PMLR},
pdf_ = {https://raw.githubusercontent.com/mlresearch/v270/main/assets/kim25c/kim25c.pdf},
url = {https://proceedings.mlr.press/v270/kim25c.html}
}

@inproceedings{baechler2024screenai,
author = {Baechler, Gilles and Sunkara, Srinivas and Wang, Maria and Zubach, Fedir and Mansoor, Hassan and Etter, Vincent and C\u{a}rbune, Victor and Lin, Jason and Chen, Jindong and Sharma, Abhanshu},
title = {{ScreenAI}: a vision-language model for {UI} and infographics understanding},
year = {2024},
isbn = {978-1-956792-04-1},
url = {https://doi.org/10.24963/ijcai.2024/339},
doi = {10.24963/ijcai.2024/339},
booktitle = {Proceedings of the Thirty-Third International Joint Conference on Artificial Intelligence},
articleno = {339},
numpages = {11},
location = {Jeju, Korea},
series = {IJCAI '24}
}

@article{hartsock2024vision,
author = {Hartsock, Iryna and Rasool, Ghulam},
title = {Vision-language models for medical report generation and visual questionanswering: a review},
journal = {Frontiers Artif. Intell.},
volume = {7},
year = {2024},
url = {https://doi.org/10.3389/frai.2024.1430984},
doi = {10.3389/FRAI.2024.1430984}
}

@article{kawaharazuka2025vision,
author = {Kawaharazuka, Kento and Oh, Jihoon and Yamada, Jun and Posner, Ingmar and Zhu, Yuke},
journal = {IEEE Access},
title = {Vision-Language-Action Models for Robotics: {A} Review Towards Real-World Applications},
year = {2025},
volume = {13},
number = {},
pages = {162467--162504},
doi = {10.1109/ACCESS.2025.3609980}
}

@article{zhang2024vision,
author = {Zhang, Jingyi and Huang, Jiaxing and Jin, Sheng and Lu, Shijian},
title = {Vision-Language Models for Vision Tasks: A Survey},
year = {2024},
issue_date = {Aug. 2024},
publisher = {IEEE Computer Society},
address = {USA},
volume = {46},
number = {8},
issn = {0162-8828},
url = {https://doi.org/10.1109/TPAMI.2024.3369699},
doi = {10.1109/TPAMI.2024.3369699},
journal = {IEEE Trans. Pattern Anal. Mach. Intell.},
month = aug,
pages = {5625--5644},
numpages = {20}
}

@article{tschannen2025siglip,
author = {Michael Tschannen and {Alexey A.} Gritsenko and Xiao Wang and {Muhammad Ferjad} Naeem and Ibrahim Alabdulmohsin and Nikhil Parthasarathy and Talfan Evans and Lucas Beyer and Ye Xia and Basil Mustafa and {Olivier J.} H{\'{e}}naff and Jeremiah Harmsen and Andreas Steiner and Xiaohua Zhai},
title = {{SigLIP} 2: {M}ultilingual Vision-Language Encoders with Improved Semantic Understanding, Localization, and Dense Features},
journal = {CoRR},
volume = {abs/2502.14786},
year = {2025},
url = {https://doi.org/10.48550/arXiv.2502.14786},
doi = {10.48550/ARXIV.2502.14786},
eprinttype = {arXiv},
eprint = {2502.14786}
}

@article{oracleProblemTSE2014,
author = {Barr, Earl T. and Harman, Mark and McMinn, Phil and Shahbaz, Muzammil and Shin Yoo},
title = {The Oracle Problem in Software Testing: A Survey},
year = {2015},
issue_date = {May 2015},
publisher = {IEEE Press},
volume = {41},
number = {5},
issn = {0098-5589},
url_ = {https://doi.org/10.1109/TSE.2014.2372785},
doi = {10.1109/TSE.2014.2372785},
journal = {IEEE Trans. Softw. Eng.},
month = may,
pages = {507--525},
numpages = {19}
}

@article{Sahoo2024PromptEngineeringSurvey,
author = {Sahoo, Pranab and Singh, Ayush Kumar and Saha, Sriparna and Jain, Vinija and Mondal, Samrat and Chadha, Aman},
title = {A Systematic Survey of Prompt Engineering in Large Language Models: Techniques and Applications},
journal = {CoRR},
volume = {abs/2402.07927},
year = {2024},
url = {https://doi.org/10.48550/arXiv.2402.07927},
doi = {10.48550/ARXIV.2402.07927},
eprinttype = {arXiv},
eprint = {2402.07927}
}

@inproceedings{ArcuriICSE11,
author = {Arcuri, Andrea and Briand, Lionel},
title = {A Practical Guide for Using Statistical Tests to Assess Randomized Algorithms in Software Engineering},
year = {2011},
isbn = {9781450304450},
publisher = {Association for Computing Machinery},
address = {New York, NY, USA},
url = {https://doi.org/10.1145/1985793.1985795},
doi = {10.1145/1985793.1985795},
booktitle = {Proceedings of the 33rd International Conference on Software Engineering},
pages = {1--10},
numpages = {10},
location = {Waikiki, Honolulu, HI, USA},
series = {ICSE '11}
}

@article{cosmosNvidiaArxiv,
author = {Niket Agarwal and Arslan Ali and Maciej Bala and Yogesh Balaji and Erik Barker and Tiffany Cai and Prithvijit Chattopadhyay and Yongxin Chen and Yin Cui and Yifan Ding and Daniel Dworakowski and Jiaojiao Fan and Michele Fenzi and Francesco Ferroni and Sanja Fidler and Dieter Fox and Songwei Ge and Yunhao Ge and Jinwei Gu and Siddharth Gururani and Ethan He and Jiahui Huang and Jacob Samuel Huffman and Pooya Jannaty and Jingyi Jin and Seung Wook Kim and Gergely Kl{\'{a}}r and Grace Lam and Shiyi Lan and Laura Leal{-}Taix{\'{e}} and Anqi Li and Zhaoshuo Li and Chen{-}Hsuan Lin and Tsung{-}Yi Lin and Huan Ling and Ming{-}Yu Liu and Xian Liu and Alice Luo and Qianli Ma and Hanzi Mao and Kaichun Mo and Arsalan Mousavian and Seungjun Nah and Sriharsha Niverty and David Page and Despoina Paschalidou and Zeeshan Patel and Lindsey Pavao and Morteza Ramezanali and Fitsum Reda and Xiaowei Ren and Vasanth Rao Naik Sabavat and Ed Schmerling and Stella Shi and Bartosz Stefaniak and Shitao Tang and Lyne Tchapmi and Przemek Tredak and Wei{-}Cheng Tseng and Jibin Varghese and Hao Wang and Haoxiang Wang and Heng Wang and Ting{-}Chun Wang and Fangyin Wei and Xinyue Wei and Jay Zhangjie Wu and Jiashu Xu and Wei Yang and Yen{-}Chen Lin and Xiaohui Zeng and Yu Zeng and Jing Zhang and Qinsheng Zhang and Yuxuan Zhang and Qingqing Zhao and Artur Z{\'{o}}lkowski},
title = {Cosmos World Foundation Model Platform for Physical {AI}},
journal = {CoRR},
volume = {abs/2501.03575},
year = {2025},
url = {https://doi.org/10.48550/arXiv.2501.03575},
doi = {10.48550/ARXIV.2501.03575},
eprinttype = {arXiv},
eprint = {2501.03575}
}

@misc{SupplementaryMaterialAIWARE2026,
author = {Saurabh, Prasun and Valle, Pablo and Arrieta, Aitor and Ali, Shaukat and Arcaini, Paolo},
year = {2026},
title = {Supplementary material for the paper {``{\approach: A Vision-Language Model-based Test Oracle for Testing Robots}''}},
url = {https://doi.org/10.5281/zenodo.19680486}
}

@inproceedings{Chen_2024_CVPR,
author = {Chen, Boyuan and Xu, Zhuo and Kirmani, Sean and Ichter, Brian and Sadigh, Dorsa and Guibas, Leonidas and Xia, Fei},
booktitle = {2024 IEEE/CVF Conference on Computer Vision and Pattern Recognition (CVPR)}, 
title = {{SpatialVLM}: Endowing Vision-Language Models with Spatial Reasoning Capabilities}, 
year = {2024},
volume = {},
number = {},
pages = {14455--14465},
doi = {10.1109/CVPR52733.2024.01370}
}

@inproceedings{chen2023autotamp,
author = {Chen, Yongchao and Arkin, Jacob and Dawson, Charles and Zhang, Yang and Roy, Nicholas and Fan, Chuchu},
booktitle = {2024 IEEE International Conference on Robotics and Automation (ICRA)}, 
title = {{AutoTAMP}: Autoregressive Task and Motion Planning with LLMs as Translators and Checkers}, 
year = {2024},
volume = {},
number = {},
pages = {6695--6702},
doi = {10.1109/ICRA57147.2024.10611163}
}

@article{VLMSocialNav,
author = {Song, Daeun and Liang, Jing and Payandeh, Amirreza and Raj, Amir Hossain and Xiao, Xuesu and Manocha, Dinesh},
journal = {IEEE Robotics and Automation Letters}, 
title = {{VLM-Social-Nav}: Socially Aware Robot Navigation Through Scoring Using Vision-Language Models}, 
year = {2025},
volume = {10},
number = {1},
pages = {508--515},
doi = {10.1109/LRA.2024.3511409}
}

@book{bishop2006,
author = {Christopher M. Bishop},
title = {Pattern Recognition and Machine Learning},
publisher = {Springer},
year = {2006}
}

@inproceedings{feng2020deepgini,
author = {Feng, Yang and Shi, Qingkai and Gao, Xinyu and Wan, Jun and Fang, Chunrong and Chen, Zhenyu},
title = {{DeepGini}: prioritizing massive tests to enhance the robustness of deep neural networks},
year = {2020},
isbn = {9781450380089},
publisher = {Association for Computing Machinery},
address = {New York, NY, USA},
url = {https://doi.org/10.1145/3395363.3397357},
doi = {10.1145/3395363.3397357},
booktitle = {Proceedings of the 29th ACM SIGSOFT International Symposium on Software Testing and Analysis},
pages = {177--188},
numpages = {12},
location = {Virtual Event, USA},
series = {ISSTA 2020}
}
